\documentclass[12pt]{article}
\usepackage{amsmath}
\usepackage{graphicx}
\usepackage{url} 
\usepackage{mwe}
\usepackage{soul}
\usepackage{comment}
\usepackage{algorithm}
\usepackage{enumerate}
\usepackage{algcompatible}
\usepackage{tikz}
\usepackage[font=small]{caption}
\usepackage{caption}
\usepackage{amssymb}
\usepackage{capt-of}
\usepackage{caption}
\usepackage{natbib}
\usepackage{bm}
\usepackage[normalem]{ulem}
\newcommand\norm[1]{\left\lVert#1\right\rVert}
\tikzset{cross/.style={cross out, draw=black, minimum size=2*(#1-\pgflinewidth), inner sep=0pt, outer sep=0pt},
cross/.default={1pt}}
\DeclareMathOperator*{\argmax}{arg\,max}
\providecommand{\abs}[1]{\lvert#1\rvert}
\renewcommand{\d}[1]{\ensuremath{\operatorname{d}\!{#1}}}

\newcommand{\blind}{0}

\newcommand{\beginsupplement}{%
        \setcounter{table}{0}
        \renewcommand{\thetable}{S\arabic{table}}%
        \setcounter{section}{0}%
        \renewcommand{\thesection}{S\arabic{section}}
        \setcounter{subsection}{0}%
        \renewcommand{\thesubsection}{S\arabic{subsection}}
        \setcounter{figure}{0}
        \renewcommand{\thefigure}{S\arabic{figure}}%
        \setcounter{equation}{0}%
        \renewcommand{\theequation}{S\arabic{equation}}
     }

\begin{document}

\def\spacingset#1{\renewcommand{\baselinestretch}%
{#1}\small\normalsize} \spacingset{1}


\if0\blind
{
  \title{\bf Neural Likelihood Surfaces for Spatial Processes with Computationally Intensive or Intractable Likelihoods}
  \author{Julia Walchessen
    \hspace{.2cm}\\
    Department of Statistics and Data Science, Carnegie Mellon University\\
    and \\
    Amanda Lenzi\\
    School of Mathematics,
    University of Edinburgh\\
    and \\
    Mikael Kuusela\\
    Department of Statistics and Data Science, Carnegie Mellon University
    }
  \date{}
  \maketitle
} \fi

\if1\blind
{
  \bigskip
  \bigskip
  \bigskip
  \begin{center}
    {\LARGE\bf Title}
\end{center}
  \medskip
} \fi

\bigskip
\begin{abstract}
In spatial statistics, fast and accurate parameter estimation, coupled with a reliable means of uncertainty quantification, can be challenging when fitting a spatial process to real-world data because the likelihood function might be slow to evaluate or wholly intractable. In this work, we propose using convolutional neural networks to learn the likelihood function of a spatial process. Through a specifically designed classification task, our neural network implicitly learns the likelihood function, even in situations where the exact likelihood is not explicitly available. Once trained on the classification task, our neural network is calibrated using Platt scaling which improves the accuracy of the neural likelihood surfaces. To demonstrate our approach, we compare neural likelihood surfaces and the resulting maximum likelihood estimates and approximate confidence regions with the equivalent for exact or approximate likelihood for two different spatial processes---a Gaussian process and a Brown--Resnick process which have computationally intensive and intractable likelihoods, respectively. We conclude that our method provides fast and accurate parameter estimation with a reliable method of uncertainty quantification in situations where standard methods are either undesirably slow or inaccurate. The method is applicable to any spatial process on a grid from which fast simulations are available.
\end{abstract}

\noindent%
\vfill

\newpage
\spacingset{1.5}

\section{Introduction}
\label{sec:intro}

In spatial statistics, parametric spatial models used to describe real-world phenomena typically have intractable or computationally intensive likelihood functions. This is generally due to the large number of spatial locations the spatial model must cover. Classical methods of statistical inference rely strongly on the likelihood function to provide parameter estimation, hypothesis testing, and uncertainty quantification \citep{Berg}. Due to this reliance, past research in the field of spatial statistics has focused on providing approximations for the likelihood that sidestep the need to evaluate the full likelihood \citep{Sun}. Examples include composite likelihood, low-rank approximations, and Vecchia approximation \citep{padoan2010likelihood, Sun}. Yet, these approximations suffer in terms of the accuracy of parameter estimation and uncertainty quantification compared to the equivalent for exact likelihood \citep{Sun}.

Thanks to the advent of modern machine learning and the ability to quickly simulate from many spatial processes, recent research has focused on neural estimation---efficient parameter estimation using neural networks---for these processes. To address the computational inefficiency of Gaussian process parameter estimation, \citet{Gerber} trained a neural network to predict the parameters of the covariance function and achieved similar accuracy yet significant computational efficiency when compared to directly computing the maximum likelihood estimator (MLE). \citet{Lenzi} demonstrated that neural estimation can be extended to other spatial processes by training a neural network to predict the parameters of several max-stable processes. 

\citet{Lenzi, SainsburyDale_2023} also proposed an uncertainty quantification method for neural estimation via bootstrapping, a computationally intensive approach with unclear theoretical and practical properties in this context. Unfortunately, neural estimation does not easily lend itself to more traditional, better-understood methods of quantifying uncertainty. Another limitation of neural estimation is the reliance on a prior over the parameter space. In this context, the prior corresponds to the distribution that was used to simulate the parameters for training the neural network. Depending on the prior selected, the neural network may produce biased estimates and unreliable bootstrapped uncertainty quantification. The prior dependence of neural estimation can be established with a simple argument: Since the neural estimator $\hat{\bm{\theta}}$ is the minimizer of the mean squared error, it can be understood as a regularized finite-sample estimator of the conditional expectation $\mathbb{E}(\bm{\theta} \mid \bm{y})$ with respect to the conditional distribution $p(\bm{\theta} \mid \bm{y}) \propto p(\bm{y} \mid \bm{\theta})p(\bm{\theta})$ which inescapably depends on the prior $p(\bm{\theta})$. Finding a way to learn the actual likelihood function instead of point predictions has the potential to address both of these limitations of neural estimation.

\begin{figure}[!t]
\centering
\begin{tikzpicture}[every text node part/.style={align=center}]
\tikzstyle{nn}=[draw,rectangle, rounded corners, fill=blue!20,
minimum width=3cm, minimum height=.75cm]
\tikzstyle{input}=[draw,rectangle, rounded corners, fill=yellow!20,
minimum width=3cm, minimum height=.75cm]
\tikzstyle{output}=[draw,rectangle, rounded corners, fill=green!20,
minimum width=3cm, minimum height=.75cm]
\tikzstyle{a}=[very thick,->,>=stealth]
\node[input] (y1) at (-2,0) {spatial field\\ $\textbf{y}\in \mathbb{R}^{s\times s}$};
\node[input] (y2) at (2.5,0) {spatial field\\ $\textbf{y}\in \mathbb{R}^{s\times s}$};
\node[input] (theta) at (6.25,0) {parameter $\bm{\theta}\in \mathbb{R}^{k}$};
\node[nn] (NN1) at (-2,-1.5) {NN};
\node[nn] (NN2) at (4,-1.5) {NN};
\node[output] (output1) at (-2,-3) {Parameter Point\\ Estimator $\hat{\bm{\theta}}$};
\node[output] (output2) at (4,-2.75) {$\propto \mathcal{L}(\bm{\theta} \mid \bm{y})$};
\node[output] (output3) at (4,-4)
{Neural Likelihood Surface};
\node[output] (output4) at (8.5,-1.5)
{Parameter Point\\ Estimator  $\hat{\bm{\theta}}$};
\node[output] (output5) at (8.5,-3.5)
{Approximate\\Confidence\\Region $\mathcal{C}_{1-\alpha}$};
\draw[-to,line width=4pt] (y1) -- (NN1);
\draw[-to,line width=4pt] (NN1) -- (output1);
\draw[-to,line width=3pt] (y2) -- (NN2);
\draw[-to,line width=3pt] (theta) -- (NN2);
\draw[-to,line width=4pt] (NN2) -- (output2);
\draw[-to,line width=4pt] (output2) -- (output3);
\draw[-to,line width=4pt] (6.35,-3.65) -- (7,-2.05);
\draw[-to,line width=4pt] (output3) -- (output5);
\end{tikzpicture}
\caption{The basic structure for traditional neural estimation (left) and learning the likelihood via our proposed method (right) where NN refers to neural network.}
\label{fig:neuralestimationlikelihood}
\end{figure}
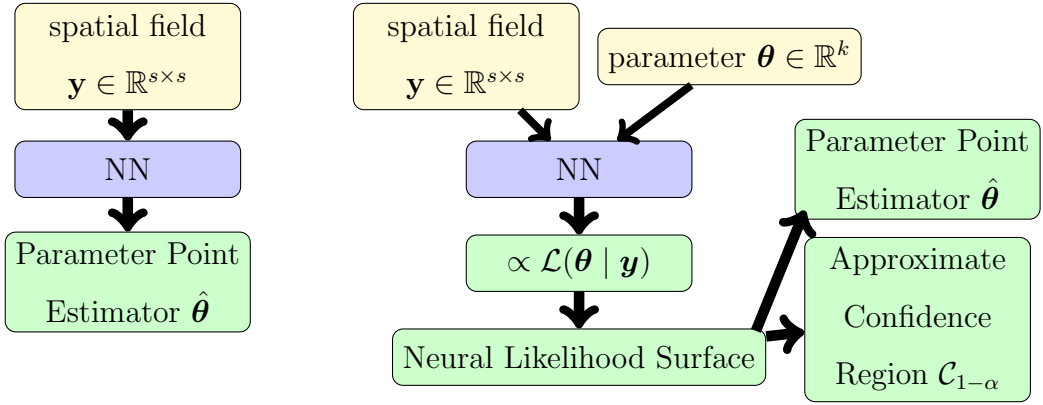
In this paper, we propose a method to learn the likelihood functions of spatial processes for which fast simulation is possible using a specifically constructed binary classification task and apply this method to processes with intractable or computationally intensive likelihoods. For the selected spatial process, fast simulation is necessary because we use simulations to form two classes consisting of pairs of a parameter $\bm{\theta}$ and a spatial field realization $\bm{y}$. The parameter $\bm{\theta}$ and spatial field $\bm{y}$ are dependent in the first class and independent in the second class. Yet, the marginal distributions for $\bm{y}$ and $\bm{\theta}$ in both classes are the same due to a permutation trick in which we form the data for the second class from the first class by permuting the parameters assigned to the realizations in the first class. We train a classifier to discriminate between these two classes and show that the resulting classifier is related to the likelihood of the spatial process via a closed-form transformation. Using the classifier and this transformation, we can produce learned likelihood surfaces, parameter estimates, and confidence regions. This method of learning the likelihood via classification is well-known in the field of simulation-based inference \citep{Cranmer2016, Hermans, Dalmasso2020}. The novelty of the present work is in introducing this method to spatial statistics and demonstrating its benefits in that context.

Since convolutional neural networks (CNNs, \citet{Lecun}) have excellent empirical performance in tasks involving image data, we use a CNN as our classifier. For neural estimation, \citet{Lenzi, Gerber, SainsburyDale_2023, Lenzi_2023, Richards_2023} used CNNs as well. However, a key distinction between neural estimation and our method is the inputs for the CNN--a single input, a spatial field $\bm{y}$, for neural estimation and two inputs, a spatial field $\bm{y}$ and parameter $\bm{\theta}$, for our method--and the outputs of the CNN--a point estimator $\bm{\hat{\theta}}$ for neural estimation and a scalar which is proportional to the likelihood up to a known transformation for our method. See Figure~\ref{fig:neuralestimationlikelihood} to better understand the basic differences between neural estimation and our method.

Learning the likelihood accurately requires our CNN to be well calibrated (i.e., to produce accurate class probabilities) so we apply post-hoc calibration to it using Platt scaling \citep{Guo}. Calibration significantly improves the predicted probabilities from the CNN and hence the learned likelihood. As such, calibration is an integral part of this method. We refer to the likelihood function that we compute based on the final trained and calibrated CNN as neural likelihood.

As a proof of concept, we first apply our method to Gaussian processes, the most popular stochastic process for modeling spatial data, and show that the neural likelihood is comparable to the exact likelihood in terms of parameter estimation and uncertainty quantification with a significant speedup in computation. Since this method of learning the likelihood is primarily intended for spatial processes with intractable likelihoods, we next test our method on a Brown--Resnick process, a model for spatial extremes with an intractable likelihood even for a moderate number of spatial locations, and demonstrate that neural likelihood is comparable or better than standard methods for approximating this likelihood in terms of parameter estimation, uncertainty quantification, and computational efficiency.

A major benefit of neural likelihood over prediction-based approaches is that it is independent of the prior over the parameter space used to generate training data for the neural network. As mentioned above, prediction-based estimators have the potential to suffer from bias depending on the selected prior over the parameter space. For our proposed method, the prior over the parameter space only affects how quickly or to what detail the neural network implicitly learns the likelihood not whether bias is introduced to the neural likelihood.

Our proposed method produces parameter estimates grounded in classical statistical inference. To produce parameter estimates, we evaluate the neural likelihood on a fine grid over the parameter space and find the maximum likelihood estimator (MLE) over this grid. Neural likelihood is particularly well suited for grid evaluations like this in terms of computational efficiency. Since we use grid search, our method is likely to find the global maximizer even in the case of a non-concave likelihood surface whereas other gradient-based methods might not.

By learning the likelihood, we can provide a means of uncertainty quantification using approximate confidence regions derived from the shape of the likelihood surface. These confidence regions are constructed using likelihood ratios as shown in Section~\ref{sec:approxConfReg}. While these confidence regions only provide approximate guarantees for finite spatial domains, empirically we find well-calibrated coverage for these regions for both Gaussian and Brown--Resnick process parameters.

Finally, the last two benefits of our method are computational efficiency and the ability to easily handle multiple realizations. First, compared to traditional methods of computing the exact or approximate likelihood, our method is significantly faster by potentially several orders of magnitude depending on the spatial process in question and the number of observed spatial locations because neural networks consist of simple non-linear and linear functions composed together. In spatial statistics, a long line of research has focused on methods to speed up likelihood evaluation for tractable spatial processes, especially the Gaussian process. Yet, as explained in Section \ref{sec:backgroundspatial}, these methods can be efficient but are often inaccurate due to loss of information. In contrast, our method can provide significant computational speed-ups with accuracy that remains comparable to exact likelihood. Lastly, since the likelihood of multiple independent realizations is the product of the single-realization likelihoods, we can handle an arbitrary number of realizations without retraining the neural network unlike the estimators proposed in \citet{Lenzi, Gerber}. Although, \citet{SainsburyDale_2023} partially addresses the multiple independent realizations case for neural estimation with a permutation-invariant neural network.

The structure of the rest of this paper is as follows. Section~\ref{sec:background} details how simulation-based inference has evolved with the aid of modern machine learning and how the field of spatial statistics has previously addressed inference for spatial processes with intractable or computationally intensive likelihoods. Section~\ref{sec:methods} describes our method of learning the likelihood including how the classifier relates to the likelihood, how the training data for the classification task is generated using simulation, why we use a neural network as the classifier, and how to produce neural likelihood surfaces, parameter estimates, and approximate confidence regions from the classifier. In Section~\ref{sec:casestudies}, we study our method on two spatial processes--Gaussian and Brown--Resnick processes--and compare the resulting neural likelihood surfaces, parameter estimates, and confidence regions to those resulting from conventional methods. Finally in Section \ref{sec:discuss}, we discuss the limitations of our method and possible future extensions.

\section{Background and Related Work}
\label{sec:background}
Thanks to advances in computing power and data storage, there is great interest in the collection, storage, and analysis of increasingly large spatial datasets. As a result, recent developments in the field of spatial statistics have focused on adapting classical statistical methods such as likelihood inference for large spatial data \citep{Sun}. In conjunction, the field of simulation-based inference (SBI) has seen tremendous development in methods for learning surrogate likelihoods, likelihood ratios, and posteriors for high-dimensional data with the help of machine learning \citep{Cramner}. In this section, we briefly describe relevant developments in both spatial statistics and SBI which provide the backdrop for this work.

\subsection{Simulation-Based Inference}
\label{sec:backgroundsbi}
In simulation-based inference, there are two classical approaches. The first is Approximate Bayesian Computation (ABC) in which simulated and observed data are compared using a distance metric and typically a summary statistic to obtain samples from an approximate posterior \citep{Sisson}. The second approach is density estimation (approximate frequentist computation) in which the distribution of the simulated data or summary statistics of the simulated data is approximated using traditional density estimation methods \citep{Cramner}. Both methods suffer from the curse of dimensionality: in some cases, the number of simulations necessary for sufficient approximation grows exponentially with the dimension. The efficiency of both classical approaches to SBI can be improved by utilizing machine learning which can more easily handle high-dimensional data \citep{Cramner}.

Beyond improving traditional approaches of SBI, machine learning has ushered in a new approach in which a surrogate for the likelihood or likelihood ratio is learned using training data generated from the simulator \citep{Cranmer2016, Dalmasso2020, Dalmasso, Cramnerparticlephysics, Brehmer2018MiningGF}. Our proposed method is directly derived from this approach of learning a likelihood surrogate via training a neural network on simulated data. As far as we are aware, there are three main approaches to learning the surrogate using machine learning \citep{Cramnerparticlephysics}. The first approach is neural density estimators which are flexible probabilistic models with tractable likelihoods that can be combined in such a way as to provide a sufficient surrogate for the likelihood. This approach differs from our proposed method by additionally approximating the normalizing constant $p(\bm{y})$ where $\bm{y}$ is the observed data. Since likelihood-based parameter estimation and uncertainty quantification do not depend on the normalizing constant $p(\bm{y})$, our proposed method is potentially much faster and less tricky to train than neural density estimation.

The second approach is CARL, a method of learning the likelihood ratio via classification and a transformation utilizing the likelihood ratio trick \citep{Cranmer2016}. Our proposed method is derived from this approach and shares some key similarities. The first similarity between CARL and our proposed method is the classifier output transformation to produce the likelihood ratio. The second similarity lies in the classification task. Yet, the classification task in CARL relies on a reference distribution whereas ours does not. Additionally, CARL utilizes a different calibration method which may significantly impact the resulting learned likelihood. 

Subsequent work inspired by CARL includes \citet{Kaji2023} and \citet{Hermans} in which a similar process of learning the likelihood ratio is employed to facilitate quick computation in Markov chain Monte Carlo algorithms. Additionally, \citet{Dalmasso2020, Dalmasso} use the learned likelihood ratio in test inversion to obtain confidence sets and \citet{Rizvi2024} analyze how the loss function and classifier structure affect the accuracy of the learned likelihood ratio.

Finally, the third approach involves incorporating knowledge about the inner workings of the simulator, such as latent variables, into learning the surrogate \citep{Brehmer2018MiningGF, Cramnerparticlephysics}. This third approach is most ideal for physical models in which there are typically latent variables and some knowledge of the simulator. This approach could also be helpful for simulation-based inference for hierarchical statistical models. However, in the absence of a hierarchical model and in the interest of full generality, we will focus in this paper on simulation-based inference that does not attempt to make explicit use of latent variables.

To the best of our knowledge, simulation-based inference for learning the likelihood has not been previously applied to spatial statistics. As such, we believe that our approach is novel in the field of spatial statistics, yet it is well-grounded in recent developments within the field of simulation-based inference, especially in high-energy physics.

\subsection{Spatial Statistics}
\label{sec:backgroundspatial}

In spatial statistics, work on likelihood-based inference methods focuses on either expediting the process of evaluating the exact likelihood for spatial processes with tractable likelihoods or providing an approximation of the likelihood for spatial processes with intractable likelihoods such as composite likelihood 
\citep{Varin2011}.

Often, when evaluating exact likelihoods of spatial processes over a large number of spatial locations, the covariance matrix must be inverted, as is the case for Gaussian processes \citep{Sun}. The time complexity and memory required for matrix inversion are $\mathcal{O}(n^{3})$ and $\mathcal{O}(n^{2})$, respectively, where $n$ is the number of spatial locations \citep[p. 56]{Sun}. To reduce the time complexity of matrix inversion, the original covariance matrix is replaced with either a low-rank or sparse matrix \citep{Heaton}. Low-rank and sparse matrices are much faster to invert than the original covariance matrix. Yet, these representations of the original covariance matrix contain less information about the spatial process which can lead to reduced accuracy in parameter estimation and uncertainty quantification and a lack of statistical guarantees \citep{Sun}.

For Gaussian random fields (GRFs), \citet{lindgren2011explicit} modeled the spatial field as a solution of an Stochastic Partial Differential Equation (SPDE) and showed that inference is of the order $\mathcal{O}(n^{3/2})$ compared to the $\mathcal{O}(n^{3})$ complexity for estimating covariance functions.
These computational benefits are due to estimating the generally sparse precision matrices (inverse covariance matrices) rather than dense covariance matrices.
However, there are several important classes of spatial processes, such as non-Gaussian models for spatial extremes, that cannot be handled using the SPDE approach.

Another popular method for efficiently evaluating the likelihood of a Gaussian process is Vecchia approximation---a method to approximate the joint density of a spatial process observed at multiple locations as a product of conditional densities according to some ordering of a subset of locations \citep{Sun,Katzfuss}.
Due to the decision to reduce the number of conditioned spatial locations, Vecchia approximation loses information about the full likelihood and requires design choices as to how to order the spatial locations and which spatial locations to not include in the conditioning.

While we cover composite likelihood for Brown--Resnick processes in greater depth in Section~\ref{sec:casestudies}, we provide a general description of composite likelihood here. Composite likelihood \citep{Varin2011} is applicable to many max-stable processes for which the likelihood is tractable only up to a certain number of spatial locations \citep{Castruccio}. For these spatial processes, the full likelihood involves a summation indexed by the number of partitions for $n$ spatial locations which grows more than exponentially with respect to $n$. Due to this more than exponential growth, the full likelihood is intractable for large and even moderate $n$. Yet, if we approximate the likelihood with a product of $m-$variate densities for $m<<n$, such as the bivariate density, we reduce the number of terms to consider. However, even this reduction in terms can be too intensive to compute, and consequentially, only a subset of spatial locations are typically used in the $m-$variate densities \citep{padoan2010likelihood}. As with Vecchia approximation, composite likelihood contains less information about the full likelihood and requires a design choice in terms of which subsets of spatial locations to include. A poor design choice may lead to a significant deterioration in the quality of inference as shown in Section \ref{supplement:additionalresults} of the Supplementary Material for the Brown--Resnick~process.

In all the techniques outlined in this (non-exhaustive) overview of methods for evaluating or approximating exact likelihoods for large spatial data, there are common themes in terms of loss of information and nontrivial design choices. Our proposed method learns a surrogate of the exact likelihood that becomes increasingly accurate the more training data is available and only requires design choices in terms of the neural network structure and training hyperparameters, choices that can be optimized using a validation dataset. Finally, many of these previous techniques are specific to certain types of spatial processes, whereas our method of learning the likelihood is generalizable to any spatial process on a grid (in fact, any statistical model with regular outputs) for which fast simulation is possible.

\section{Methodology}
\label{sec:methods}
This section describes our framework for learning the likelihood function from a specifically-designed classification task using simulated data from a given spatial process. Our proposed methodology is derived from previous work in simulation-based inference on learning the likelihood ratio via classification \citep{Cranmer2016, Hermans, Dalmasso2020}. Specifically, our task is to infer the parameter $\bm{\theta}$ from a realization $\bm{y}$ of a spatial process evaluated on a finite set of spatial locations:
\begin{equation}
\begin{aligned}
&f \sim \textrm{SpatialProcess}(\bm{\theta}),\\
&\bm{y} := f(\mathcal{S}),
\end{aligned}
\label{eqn:inferenceprob}
\end{equation} where $\mathcal{S}$ is a set of spatial locations on which we observe a realization $f$ from a spatial process depending on an unknown parameter $\bm{\theta}$. A key assumption throughout this work is that for any $\bm{\theta}$, it is easy to simulate realizations $\bm{y}$ from $\textrm{SpatialProcess}(\bm{\theta})$ according to Equation~\eqref{eqn:inferenceprob}.

\subsection{Connection between Simulated Data, Classification Task, and Likelihood}
\label{sec:methodslikelihoodconnection}
Our methodology hinges on four key observations. First, simulation from many spatial processes with intractable likelihoods is simple and fast. Fast simulation enables us to apply likelihood-free inference techniques to these intractable spatial processes in order to learn the likelihood \citep{Cramner}. Second, the realizations from spatial processes with different parameters may be distinguishable by the well-trained eye, and as such, one would expect an image classification model such as a convolutional neural network (CNN) to be able to distinguish between realizations generated by different parameters. Third, modern classification algorithms are not necessarily affected by the curse of dimensionality to the same extent as other methods of directly learning the likelihood might be. Finally, following \citet{Cranmer2016, Hermans, Dalmasso2020}, we can draw the connection below between a specifically-designed classification task and the likelihood function of a given spatial process using a permutation-based approach described in Section~\ref{sec:methodclasses}.

Consider the space $D\times \Theta$, where $D$ is the space of realizations for some spatial process evaluated on spatial locations $\mathcal{S}$ governed by parameters from the parameter space $\Theta$ which we assume to be bounded. We will relate the likelihood function $\mathcal{L}(\bm{\theta} \mid \bm{y})=p(\bm{y}\mid \bm{\theta})$ for parameter $\bm{\theta} \in \Theta$ and realization $\bm{y}\in D$ to the output of a probabilistic binary classifier $h:D\times \Theta \to [0,1], (\bm{y},\bm{\theta}) \mapsto h(\bm{y},\bm{\theta})$. The first class $(C=1)$ for the binary classifier is the class in which pairs of realizations and parameters are dependent. Specifically, the given realization $\bm{y}$ is generated from the spatial process with the given parameter $\bm{\theta}$. The second class $(C=2)$ is the class in which the pairs of realizations and parameters are independent yet have the same marginal distributions as in the first class. The conditional probabilities for the two classes have the following form:
\begin{equation}
P\big((\bm{y},\bm{\theta}) \mid C = 1\big) = p(\bm{y},\bm{\theta}) \quad\textrm{and}\quad
P\big( (\bm{y},\bm{\theta})\mid C = 2\big) = p(\bm{y})p(\bm{\theta}),
\label{eqn:conditionalprobs}
\end{equation} where $p(\bm{y})$ and $p(\bm{\theta})$ are the marginal distributions implied by $p(\bm{y},\bm{\theta})$,
\begin{equation}
p(\bm{y}) = \int p(\bm{y}, \bm{\theta})\, \mathrm{d} \bm{\theta} \quad\textrm{and}\quad
p(\bm{\theta}) = \int p(\bm{y},\bm{\theta}) \, \mathrm{d} \bm{y}.
\label{eqn:marginaldist}
\end{equation}
We can now draw the connection between our constructed classifier $h(\bm{y},\bm{\theta})$ and the likelihood function $\mathcal{L}(\bm{\theta} \mid \bm{y})$:
\begin{equation}
\begin{aligned}
h(\bm{y},\bm{\theta}) & = P\big(C = 1 \mid (\bm{y},\bm{\theta}) \big)
= \frac{P\big((\bm{y},\bm{\theta})\mid C=1\big)P(C=1)}{P\big( (\bm{y},\bm{\theta}) \mid C=1\big)P(C=1) + P\big( (\bm{y},\bm{\theta})\mid C= 2\big)P(C=2)}\\
& = \frac{\frac{P\big( (\bm{y},\bm{\theta}) \mid C=1\big)}{P\big( (\bm{y},\bm{\theta})\mid C=2\big)}}{\frac{P\big((\bm{y},\bm{\theta})\mid C=1\big)}{P\big( (\bm{y},\bm{\theta}) \mid C=2\big)}+1}
= \frac{\frac{p(\bm{y},\bm{\theta})}{p(\bm{y})p(\bm{\theta})}}{\frac{p(\bm{y},\bm{\theta})}{p(\bm{y})p(\bm{\theta})} + 1}= \frac{p(\bm{y})^{-1}p(\bm{y} \mid \bm{\theta})}{p(\bm{y})^{-1}p(\bm{y} \mid \bm{\theta}) + 1} = \frac{p(\bm{y})^{-1}\mathcal{L}(\bm{\theta} \mid \bm{y})}{p(\bm{y})^{-1} \mathcal{L}(\bm{\theta} \mid \bm{y}) + 1}.
\end{aligned}
\label{eqn:initialclassificationconnection}
\end{equation}
On the first line of \eqref{eqn:initialclassificationconnection}, we apply Bayes' Rule in order to describe the classifier output $h(\bm{y},\bm{\theta})$ in terms of the conditional probabilities for the two classes. In the second line, we use \eqref{eqn:conditionalprobs} and assume the two classes are balanced, i.e., $P(C=1) = P(C=2) = 1/2$, for ease of exposition.

To more clearly connect the classifier and the likelihood function, we define a function $\psi: D\times \Theta \to \mathbb{R}$ in the following way:
\begin{equation}
\psi(\bm{y},\bm{\theta}) = p(\bm{y})^{-1}\mathcal{L}(\bm{\theta} \mid \bm{y}).
\label{eqn:psidef}
\end{equation}
Substituting the function $\psi$ into \eqref{eqn:initialclassificationconnection}, we obtain
\begin{equation}
h(\bm{y},\bm{\theta})=\frac{\psi(\bm{y},\bm{\theta})}{\psi(\bm{y}, \bm{\theta}) + 1}.
\label{eqn:psiclassifier}
\end{equation} We can solve \eqref{eqn:psiclassifier} for $\psi$ which is proportional to the likelihood function $\mathcal{L}(\bm{\theta} \mid \bm{y})$ for a fixed realization $\bm{y}$:
\begin{equation}
\mathcal{L}(\bm{\theta} \mid \bm{y}) \propto \psi(\bm{y},\bm{\theta})
= \frac{h(\bm{y},\bm{\theta})}{1-h(\bm{y},\bm{\theta})}.
\label{eqn:classificationconnection}
\end{equation} This shows that the likelihood function is proportional to a simple transformation of the classifier output $h$ in the case of balanced classes. The same holds for the case of unbalanced classes; the right side of \eqref{eqn:classificationconnection} is simply multiplied by the fixed ratio $\frac{P(C=2)}{P(C=1)}$.

If desired, we can also use the same classifier $h$ to evaluate the likelihood (up to a multiplicative constant) for an arbitrary number of i.i.d.\ realizations $\bm{y}_{1},\dots,\bm{y}_{n}$ from a common spatial process in the following manner:
\begin{equation}
\mathcal{L}(\bm{\theta} \mid \bm{y}_{1},\dots,\bm{y}_{n}) = \prod_{i=1}^{n} \mathcal{L}(\bm{\theta} \mid \bm{y}_{i}) \propto \prod_{i=1}^{n} \psi(\bm{y}_{i}, \bm{\theta}) = \prod_{i=1}^{n} \frac{h(\bm{y}_{i},\bm{\theta})}{1-h(\bm{y}_{i},\bm{\theta})}.
\label{eqn:multiclassificationconnection}
\end{equation}
Notice that there is no need to change the classifier $h$ if $n$ changes.

In the following section, we explain how to construct the two classes for training a classifier $\hat{h}$ which estimates the exact classifier $h$ from finite simulated data. With the estimator $\hat{h}$, we can use \eqref{eqn:psidef} and \eqref{eqn:classificationconnection} to obtain the estimated neural likelihood,
\begin{equation}
\mathcal{L}_{\mathrm{NN}}(\bm{\theta} \mid \bm{y}):=p(\bm{y})\hat{\psi}(\bm{y},\bm{\theta}) \propto \hat{\psi}(\bm{y},\bm{\theta})=\frac{\hat{h}\big( \bm{y},\bm{\theta}\big)}{1-\hat{h}\big(\bm{y},\bm{\theta}\big)}.
\label{eqn:classificationapproximationconnection}
\end{equation}

\subsection{Generating the two specifically constructed classes via simulation}
\label{sec:methodclasses}
In order to train the classifier $\hat{h}:D\times \Theta \to [0,1]$ which is an estimator of the exact classifier $h$ in \eqref{eqn:initialclassificationconnection}, we need to first obtain samples from the two classes described above. To produce the first class ($C=1$), we simulate realizations $\bm{y}$ from the spatial process for different parameters $\bm{\theta} \in \Theta$ and pair the simulated realizations with the parameter which generated the realization. To produce the second class ($C=2$), we simply permute the parameters assigned to the simulated realizations in the first class to break the dependency between the two while maintaining the marginal distributions. In the following paragraphs, we describe in detail how the classes are constructed using simulation and permutation and provide pseudocode for the construction process in Algorithms \ref{algo:firstclass} and \ref{algo:secondclass}.

First, we sample $m$ parameters from the bounded parameter space $\Theta$ in such a way that the sampled parameters are guaranteed to uniformly cover $\Theta$. It is key to note that $\Theta$ must be bounded in order to obtain samples which accurately represent the whole parameter space. It is also important to note that the sampling method for $\Theta$ is not a prior over $\Theta$ in the Bayesian sense. Our method of learning the likelihood does not require any priors over $\Theta$ in the Bayesian sense. Yet, a poor sampling method that does not provide sufficient coverage of $\Theta$ will result in a neural likelihood surface with less information about the true likelihood in certain regions of $\Theta$. To avoid this, we utilize Latin hypercube sampling (LHS) using the LHS R package \citep{lhs}. In Latin hypercube sampling, $\Theta$ is partitioned into $m$ equal sized regions, and a parameter $\bm{\theta}_{i}$ is sampled from each of these $m$ regions and forms the sampled parameters $\{\bm{\theta}\}_{i\in [m]}$. As such, Latin hypercube sampling guarantees that the sampled parameters $\{\bm{\theta}\}_{i\in [m]}$ provide uniform coverage of $\Theta$.

Next, for each sampled parameter, we simulate $n$ realizations from the spatial process evaluated at locations $\mathcal{S}$. We will specify what $\mathcal{S}$ is in Section \ref{sec:methodsarchitecture}. The $n \cdot m$ pairs of realizations and the sampled parameters which generated the realizations will form the first class. The sampling process for the first class is summarized in Algorithm~\ref{algo:firstclass}.

After simulating data for the first class, we can construct the independent pairs of realizations and parameters from the dependent pairs of the first class using permutations of the sampled parameters. Since there are $m$ sampled parameters and $n$ realizations per sampled parameter, we randomly sample $n$ permutations of the indices $1$ through $m$ which we refer to as $\pi_{1},\dots,\pi_{n}$. For the $j$th realizations of the spatial fields, we apply the permutation $\pi_{j}$ to the sampled parameters and assign each of the permuted parameters to the corresponding spatial field realization to obtain pairs of the following form: $\{(\bm{y}_{i,j}, \bm{\theta}_{\pi_{j}(i)}) \mid i\in [m]\}$.

\begin{figure}[!t]
\centering
\begin{tikzpicture}[every text node part/.style={align=center},scale = 1]
\tikzstyle{nn}=[draw,rectangle, rounded corners, fill=blue!20,
minimum width=2cm, minimum height=1cm]
\tikzstyle{input}=[draw,rectangle, rounded corners, fill=yellow!20,
minimum width=2cm, minimum height=1cm]
\tikzstyle{output}=[draw,rectangle, rounded corners, fill=green!20,
minimum width=2cm, minimum height=1cm]
\tikzstyle{a}=[very thick,->,>=stealth]
\node[input, label = {\textbf{First Class}}] (first) at (-4,0) {
$\begin{bmatrix}
(\bm{y}_{1,1}, \bm{\theta}_{1}) & \dots & (\bm{y}_{1,n}, \bm{\theta}_{1})\\
\vdots & \ddots & \vdots\\
(\bm{y}_{m,1}, \bm{\theta}_{m}) & \dots & (\bm{y}_{m,n}, \bm{\theta}_{m})
\end{bmatrix}$
};

\node[input, label = {\textbf{Second Class}}] (second) at (4,0) {
$\begin{bmatrix}
(\bm{y}_{1,1}, \bm{\theta}_{\pi_{1}(1)}) & \dots & (\bm{y}_{1,n}, \bm{\theta}_{\pi_{n}(1)})\\
\vdots & \ddots & \vdots\\
(\bm{y}_{m,1}, \bm{\theta}_{\pi_{1}(m)}) & \dots & (\bm{y}_{m,n}, \bm{\theta}_{\pi_{n}(m)})
\end{bmatrix}$
};
\end{tikzpicture}
\caption{A visualization of the pairs of spatial field realizations $\bm{y}$ and parameters $\bm{\theta}$ in the first and second classes. For both classes, the realizations $\bm{y}_{i,j}$ in the $i$th row are generated using the same $\bm{\theta}_{i}$. In the first class, all realizations $\bm{y}_{i,j}$ are paired with the $\bm{\theta}_{i}$ which generated the realization. In the second class, the realizations $\bm{y}_{i,j}$ in the $j$th column are paired with parameters permuted according to the $j$th permutation of the total number of parameters~$m$.}
\label{fig:firstsecondclass}
\end{figure}
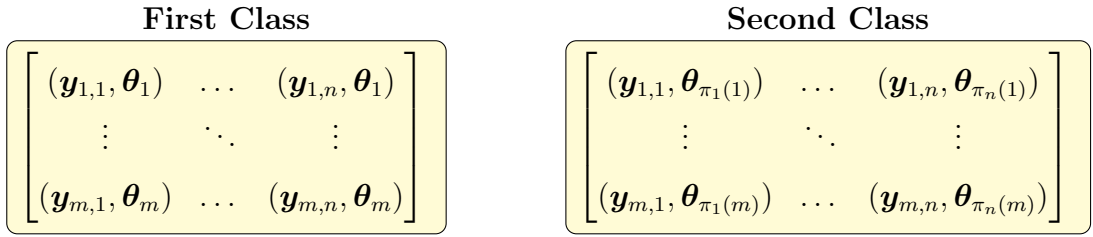

We repeat this process for all $n$ values of the index $j$. The spatial field realizations and permuted parameters are now independent of each other with unchanged marginal distributions by construction, as required to satisfy Equation~\eqref{eqn:initialclassificationconnection}. Due to this process of permuting the sampled parameters, we avoid simulating and storing new data for the second class. Pseudocode for the process of constructing the second class from the first class is shown in Algorithm~\ref{algo:secondclass}. See Figure~\ref{fig:firstsecondclass} for an illustration of the first and second classes to more fully understand the permutation process.
\begin{figure}[!t]
\spacingset{1}
\begin{minipage}{.45\textwidth}
 \begin{algorithm}[H]
 \caption{Generating Data for First Class}
 \begin{algorithmic}
  \FOR{$i$ in $1:m$}
 \STATE $\bm{\theta}_{i}\sim \textrm{LHS}(\Theta)$
    \FOR{$j$ in $1:n$}
    \STATE $f_{i,j}\sim \textrm{SpatialProcess}(\bm{\theta}_{i})$
    \STATE $\bm{y}_{i,j}=f_{i,j}(\mathcal{S})$
    \ENDFOR
 \ENDFOR
 \STATE The first class is \\$C_{1}=\{(\bm{y}_{i,j},\bm{\theta}_{i})\}_{i\in [m],j\in [n]}$.
 \end{algorithmic}
 \label{algo:firstclass}
 \end{algorithm}
 \end{minipage}%
\hfill
\begin{minipage}{0.45\textwidth}
 \begin{algorithm}[H]
 \caption{Constructing Second Class from First Class}
 \begin{algorithmic}
 \STATE Sample uniformly at random $n$ \\permutations $\pi_{1}\cdots \pi_{n}$ of indices $1\dots m$
  \STATE Initialize $C_{2} = \emptyset$
 \FOR{$j$ in $1:n$}
 \STATE $C_{2} \leftarrow C_{2} \cup \{(\bm{y}_{i,j},\bm{\theta}_{\pi_{j}(i)}) \mid i\in [m]\}$
 \ENDFOR
 \STATE The second class is \\$C_{2}=\{(\bm{y}_{i,j},\bm{\theta}_{\pi_{j}(i)})\mid i\in [m], j\in [n]\}$.
 \end{algorithmic}
 \label{algo:secondclass}
 \end{algorithm}
\end{minipage}%
\end{figure}

\subsection{Classifier Architecture}
\label{sec:methodsarchitecture}

Now that we have described how to learn the likelihood using specifically constructed classes, we need to select a classifier that can be flexible both in input shape and structure. As of yet, we have not specified the set of spatial locations $\mathcal{S}$ for the evaluation of the spatial process. In this paper, we restrict the set of spatial locations $\mathcal{S}$ to a regular grid of fixed size and discuss the potential to relax this restriction in Section \ref{sec:discuss}. The classifier needs to be flexible enough in structure to accommodate many different spatial processes since our method of learning the likelihood through classification is in principle applicable to any spatial process. 

Since neural networks are both flexible in input shape and structure and flexible in the complexity of the functions they can approximate, we use a convolutional neural network (CNN). CNNs have previously been used by \citet{Lenzi, Gerber, SainsburyDale_2023, Richards_2023} to predict parameters of spatial fields. We continue this practice here, with appropriate modifications as detailed below, because CNNs are excellent at handling images and the spatial field input $\bm{y}$ is effectively an image. See Section \ref{supplement:cnn} of the Supplementary Material for an overview of CNNs.

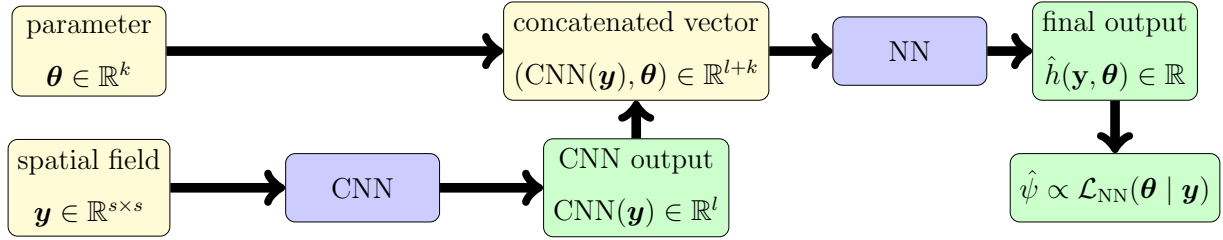
\begin{figure}[!t]
\begin{center}
\resizebox{\textwidth}{!}{%
\begin{tikzpicture}[every text node part/.style={align=center}, scale = 1]
\tikzstyle{nn}=[draw,rectangle, rounded corners, fill=blue!20,
minimum width=2.25cm, minimum height=1cm]
\tikzstyle{input}=[draw,rectangle, rounded corners, fill=yellow!20,
minimum width=2.25cm, minimum height=1cm]
\tikzstyle{output}=[draw,rectangle, rounded corners, fill=green!20,
minimum width=2.25cm, minimum height=1cm]
\tikzstyle{decision}=[draw, diamond, fill=red!20, minimum width=4cm, aspect=2]
\tikzstyle{a}=[very thick,->,>=stealth]
\node[input] (y) at (-8,-1) {spatial field\\ $\bm{y}\in \mathbb{R}^{s\times s}$};
\node[input] (theta) at (-8,1) {parameter\\ $\bm{\theta}\in \mathbb{R}^{k}$};
\node[nn] (CNN) at (-4,-1) {CNN};
\node[output] (CNN output) at (0,-1) {CNN output\\
$\textrm{CNN}(\bm{y})\in \mathbb{R}^{l}$};
\node[input] (sequential input) at (0,1) {concatenated vector\\
$(\textrm{CNN}(\bm{y}), \bm{\theta}) \in \mathbb{R}^{l+k}$};
\node[nn] (Dense NN) at (4,1) {NN};
\node[output] (final output) at (7,1) {final output\\
$\hat{h}(\textbf{y},\bm{\theta})\in \mathbb{R}$};
\node[output] (L) at (7, -1) {$\hat{\psi}\propto \mathcal{L}_{\mathrm{NN}}(\bm{\theta} \mid \bm{y})$};
\draw[-to,line width=4pt] (y) -- (CNN);
\draw[-to,line width=4pt] (CNN) -- (CNN output);
\draw[-to,line width=4pt] (CNN output) -- (sequential input);
\draw[-to,line width=4pt] (theta) -- (sequential input);
\draw[-to,line width=4pt] (sequential input) -- (Dense NN);
\draw[-to,line width=4pt] (Dense NN) -- (final output);
\draw[-to,line width=4pt] (final output) -- (L);
\end{tikzpicture}
}
\caption{The basic structure of our CNN. In this illustration, CNN refers to the convolutional and pooling layers and NN refers to the fully connected layers.}
\label{fig:basicframework}
\end{center}
\end{figure}

Since the input for our classifier is both a matrix, the spatial field $\bm{y}$, and a vector, the parameter $\bm{\theta}$, we need to modify the structure of a typical CNN to accommodate this type of input. For our CNN classifier, we first transform the spatial field into a flattened vector of reduced dimension via the convolutional part of the CNN classifier. Since the flattened vector and the parameter $\bm{\theta}$ are both in vector form, we can concatenate the two and process the concatenated vector with the fully connected part of the CNN classifier where the last layer outputs the desired class probability. The CNN classifier output is $\hat{h}(\bm{y},\bm{\theta})=\hat{P}\big(C=1 \mid (\bm{y},\bm{\theta})\big)$ which we obtain by using the sigmoid activation function and binary cross-entropy loss \citep[p.~130]{Goodfellow}. Figure~\ref{fig:basicframework}  illustrates this CNN architecture. We implement this CNN classifier using the Keras interface \citep{Keras} for Tensorflow and train the network using the Adam optimizer \citep{Kingma}.

\subsection{Calibration}

Due to growth in model complexity, modern neural networks are not always well-calibrated \citep{Guo}. A classification model is well-calibrated when the estimated class probability for a given input is sufficiently close to the true probability of the input belonging to the given class. For binary classification, perfect calibration is defined as
\begin{equation}
P(C=1 \mid \hat{h}(\bm{x})=p)=p, \quad \forall p\in [0,1],
\end{equation}
where $\bm{x}$ is the input and $\hat{h}(\bm{x})=\hat{P}(C=1\mid \bm{x})$ is the estimated probability of the input belonging to class one \citep{Guo}. For neural networks, model capacity is determined by depth, the number of layers in a neural network, and width, the complexity of each layer. Our CNN has the potential to suffer from miscalibration depending on the complexity of the convolutional layers.

To correct this potential miscalibration, we utilize a calibration method called \emph{Platt scaling} \citep{Platt}. In Platt scaling, the original class probabilities from a classifier are used as features for a logistic regression model whose response variables are the true class labels. The fitted values of this regression are the calibrated class probabilities. The calibration training data is independent from the data used to train the CNN. 
Our logistic regression model has the following form:
\begin{equation}
    \log\left(\frac{\pi(p)}{1-\pi(p)}\right) = \beta_{0}+\beta_{1}\log\left(\frac{p}{1-p}\right),
\label{eqn:plattscale}
\end{equation}
where $p$ is the uncalibrated probability for class one and $\pi(p)$ is the corresponding calibrated probability. In \eqref{eqn:plattscale}, the domain of $p$ is transformed from $[0,1]$ to $\mathbb{R}$ which improves the logistic regression model. More details on calibration are given in Section \ref{sec:casestudies}. From here on, we use $\hat{h}$ to denote the calibrated classifier.

\subsection{Likelihood Inference with Neural Likelihood}
\label{sec:methodinference}
In this section, we describe how to construct likelihood surfaces, parameter estimators, and approximate confidence regions using the neural likelihood. It is worth noting that the neural likelihood $\mathcal{L}_{\mathrm{NN}}(\bm{\theta} \mid \bm{y})=p(\bm{y})\hat{\psi}(\bm{y},\bm{\theta})$ is only known up to a multiplicative constant because $p(\bm{y})$ is unknown. In what follows, we evaluate $\hat{\psi}(\bm{y},\bm{\theta})$ over the parameter space~$\Theta$. This surface, which we refer to as the neural likelihood surface, is proportional to the likelihood $\mathcal{L}_{\mathrm{NN}}(\bm{\theta} \mid \bm{y})$. As shown below, parameter estimators and approximate confidence regions are fortunately not affected by the unknown proportionality constant.

\subsubsection{Constructing Neural Likelihood Surfaces}
\label{sec:methodinferencesurface}
First, we explain how to obtain the neural likelihood surface over the parameter space $\Theta$ from the classifier $\hat{h}$ for a fixed realization $\bm{y}\in D$. Since the parameter space $\Theta$ is bounded, we evaluate the classifier $\hat{h}(\bm{y},\cdot)$ on a regular grid
\begin{equation}
\Theta^{L} = \{(\theta_{1}+\alpha_{1} \cdot i_{1},\dots, \theta_{k}+\alpha_{k} \cdot i_{k})^T \mid i_{j}\in [s_{j}], \textrm{ }j\in [k]\}
\label{eqn:surfacegrid}
\end{equation}
where $\alpha_{1},\dots,\alpha_{k}>0$ are the fixed lengths between points for the respective dimensions, $s_{j}\in \mathbb{N}$ are the number of points in each dimension, and $\theta_{j}$ and $\theta_{j}+\alpha_{j}\cdot s_{j}$ are the endpoints of $\Theta$ for the respective dimension. Then, we transform the classifier outputs $\hat{h}(\bm{y},\bm{\theta})$ for $\bm{\theta} \in \Theta^{L}$ via the transformation given in \eqref{eqn:classificationapproximationconnection}. This produces a regular grid of values which are proportional to the neural likelihood function $\mathcal{L}_{\mathrm{NN}}(\cdot \mid \bm{y})$ evaluated on the regular grid $\Theta^{L}$ for the fixed realization $\bm{y}$.

Due to modern software packages such as the Keras \citep{Keras} interface to Tensorflow, we can quickly construct the neural likelihood surface using our CNN in the following manner. For a given realization $\bm{y}\in D$, we evaluate the CNN with the vectorized input of the realization $\bm{y}$ and parameters in the regular grid, $\{(\bm{y},\bm{\theta}_{l}) \mid \bm{\theta}_{l}\in \Theta^{L}\}$,
to obtain the neural likelihood surface. The CNN can handle this vectorized input at once, and thus, we can efficiently evaluate the neural likelihood surface. Moreover, once trained and calibrated, the CNN is amortized and thus, we can efficiently evaluate the neural likelihood surface for multiple realizations $\bm{y}$. To demonstrate our method, we compare the neural likelihood surface to the exact or approximate likelihood surface constructed using the same process on the same regular grid $\Theta_{L}$.

Finally, the surfaces we construct are truly the log neural and log exact and log approximate likelihood surfaces. As shown in Section \ref{sec:methodslikelihoodconnection}, the transformed classifier output $\hat{\psi}$ is equivalent to the neural likelihood up to a multiplicative constant. By applying a log transformation to the logit-transformed classifier output in \eqref{eqn:classificationapproximationconnection}, the multiplicative constant becomes an additive constant. Consequentially, the ranges of the log neural and log exact likelihood surfaces should have the same range of values yet shifted by some unknown constant. By constructing log neural and log exact or log approximate likelihood surfaces, we facilitate comparisons between the surfaces. 

\subsubsection{Constructing Parameter Estimators from Neural Likelihood Surfaces}
\label{sec:methodsinferenceparameter}
Parameter estimation for the neural and exact or approximate likelihoods is a discrete version of maximum likelihood estimation (MLE) in which we select the parameter on the grid $\Theta^{L}$ which maximizes the given surface,
\begin{equation}
\begin{aligned}
\hat{\bm{\theta}}_{\mathrm{NN}} = \argmax_{\bm{\theta} \in \Theta^{L}} \mathcal{L}_{\mathrm{NN}}(\bm{\theta} \mid \bm{y}) = \argmax_{\bm{\theta} \in \Theta^{L}} p(\bm{y})\hat{\psi}(\bm{y},\bm{\theta})
= \argmax_{\bm{\theta} \in \Theta^{L}} \hat{\psi}(\bm{y}, \bm{\theta}).
\end{aligned}
\label{eqn:gMLE}
\end{equation}
In the rest of the paper, we refer to the estimator in \eqref{eqn:gMLE} as the neural parameter estimator $\hat{\bm{\theta}}_{\mathrm{NN}}$. Following the same process of determining the parameter which maximizes the surface over $\Theta^{L}$, we produce exact likelihood estimators $\hat{\bm{\theta}}$ and approximate likelihood estimators $\hat{\bm{\theta}}_{\textrm{approx}}$:
\begin{equation}
\hat{\bm{\theta}} = \argmax_{\bm{\theta} \in \Theta^{L}} \mathcal{L}(\bm{\theta} \mid \bm{y})
\quad\textrm{ and }\quad 
\hat{\bm{\theta}}_{\textrm{approx}} = \argmax_{\bm{\theta} \in \Theta^{L}} \mathcal{L}_{\textrm{approx}}(\bm{\theta} \mid \bm{y}) .
\label{eqn:gMLEapproxexact}
\end{equation}

There are two major reasons why we use a grid-based approach rather than a gradient-based approach for parameter estimation via maximum likelihood. First, due to the way we construct the neural likelihood, its exact gradient is only available via automatic differentiation; see Section~\ref{sec:discuss} for further discussion. To fairly compare neural and exact or approximate likelihood, we use the same grid-based maximization approach for each function in order to produce a fair comparison between parameter estimators. Second, a gradient-based maximization approach may suffer from a local maxima problem whereas a grid-based maximization approach does not. If the exact likelihood were non-convex, a gradient-based approach might find a local maximum rather than a global maximum and hence potentially select a parameter far from the global maximizer. Yet, the grid-based approach would select a parameter which is reasonably close to the global maximizer over $\Theta$ depending on the resolution of the grid. Hence, the grid-based approach enables \emph{global maximization} of the likelihood within $\Theta$.


\subsubsection{Constructing Approximate Confidence Regions from Neural Likelihood Surfaces} \label{sec:approxConfReg}

Next, we explain how to construct approximate confidence regions from the neural, exact, and approximate likelihood surfaces. 
Specifically, we construct confidence regions from an inversion of the likelihood ratio test because these confidence regions tend to have better coverage accuracy than those constructed using inversion of other asymptotic methods \citep{Berg, Vaart}.

In the likelihood ratio test, we reject the null hypothesis that the true parameter is $\bm{\theta}_{0}$ if the $-2\times$log-likelihood ratio test statistic is greater than some critical value $c_{\alpha}$:
\begin{equation}
    \textrm{reject } H_{0} \textrm{ if } {-}2\log(\lambda(\bm{\theta}_{0})) > c_{\alpha}, \textrm{ where } \lambda(\bm{\theta}_{0})=\frac{\mathcal{L}(\bm{\theta}_{0} \mid \bm{y})}{\mathcal{L}(\hat{\bm{\theta}} \mid \bm{y})},
    \label{eqn:likelihoodratiotest}
\end{equation} with $c_{\alpha}$ chosen such that the test has significance level $\alpha$ for large enough sample sizes \citep[p. 429]{Shao}. Depending on the spatial process, the distribution of the likelihood ratio test statistic $-2\log(\lambda(\bm{\theta}))$ converges under the null hypothesis to a chi-squared distribution with degrees of freedom equal to the dimension of the parameter space $\Theta$ under increasing spatial domain asymptotics \citep[p. 50]{Gelfand}. Therefore, we set $c_{\alpha} = \chi^2_{k,1-\alpha}$, the $1-\alpha$ quantile of the $\chi^2_k$ distribution.

\citet{Mardia} showed that the maximum likelihood estimator for the covariance parameters of a Gaussian process is normally distributed with unbiased mean and the inverse of the Fisher information as the covariance matrix under increasing-domain asymptotics. As a consequence, the likelihood ratio statistic converges to a chi-squared distribution as the spatial domain increases \citep{Davison_2003}. Therefore, we expect the approximate confidence regions constructed from the likelihood ratio test to have the desired coverage in our first case study using Gaussian processes in Section~\ref{sec:GPcasestudy} because the spatial domain is sufficiently large.

Our second case study in Section~\ref{sec:casestudytwo} involves Brown--Resnick processes whose asymptotic properties are less well-studied. There is some work indicating that the asymptotic distribution of the maximum likelihood estimator in the multiple realizations case has the aforementioned properties which implies that the likelihood ratio statistic is asymptotically chi-squared distributed \citep{Dombry_2017,Davison_2003}. Consequently, if we can connect the asymptotic theory for multiple realizations with the asymptotic theory for increasing spatial domain, we have shown the likelihood ratio statistic has the desired asymptotic properties. Ergodicity provides such a connection and, indeed, under certain conditions, max-stable processes, including Brown-Resnick processes, are ergodic \citep{Stoev_2008}. As such, there is some indication that our constructed approximate confidence regions should have the intended coverage for Brown--Resnick processes, a claim that will be evaluated in detail in our experiments in Section~\ref{sec:casestudytwo}.

We can construct a $1-\alpha$ confidence set $\mathcal{C}_{1-\alpha}(\Theta)$ by inverting the likelihood ratio test \citep{Berg}. The $1-\alpha$ confidence set $\mathcal{C}_{1-\alpha}(\Theta)$ consists of all those parameters in $\Theta$ for which the null hypothesis in \eqref{eqn:likelihoodratiotest} is not rejected:
\begin{equation}
\begin{aligned}
    \mathcal{C}_{1-\alpha}(\Theta) & = \{ \bm{\theta} \in \Theta \mid -2\log(\lambda(\bm{\theta}))\leq c_{\alpha}\}\\
    & = \{\bm{\theta} \in \Theta \mid -2\Big(\log(\mathcal{L}(\bm{\theta} \mid \bm{y})) - \log(\mathcal{L}(\hat{\bm{\theta}} \mid \bm{y}))\Big) \leq c_{\alpha}\}\\
    & = \{\bm{\theta} \in \Theta \mid 2\Big(\log(\mathcal{L}(\hat{\bm{\theta}} \mid \bm{y})) - \log(\mathcal{L}( \bm{\theta} \mid \bm{y}))\Big) \leq c_{\alpha}\}.
\end{aligned}
\label{eqn:confidenceset}
\end{equation}
To accommodate the evaluation of the likelihood on a regular grid, we restrict \eqref{eqn:confidenceset} to parameters on the grid $\Theta^{L}\subset \Theta$:
\begin{equation}
    \mathcal{C}_{1-\alpha}(\Theta^{L}) = \{\bm{\theta} \in \Theta^{L} \mid 2\Big(\log(\mathcal{L}(\hat{\bm{\theta}} \mid \bm{y})) - \log(\mathcal{L}(\bm{\theta} \mid \bm{y}))\Big) \leq c_{\alpha}\}.
\label{eqn:gridconfidenceset}
\end{equation} For the approximate likelihood, we simply replace $\mathcal{L}(\bm{\theta}\mid\bm{y})$ with $\mathcal{L}_{\textrm{approx}}(\bm{\theta}\mid\bm{y})$ in \eqref{eqn:gridconfidenceset} to obtain $\mathcal{C}_{\textrm{approx},1-\alpha}(\Theta^{L})$. Finally, we can replace $\mathcal{L}(\bm{\theta}\mid \bm{y})$ with $\hat{\psi}$ to obtain $\mathcal{C}_{\mathrm{NN},1-\alpha}(\Theta^{L})$,
\begin{equation}
\begin{aligned}
\mathcal{C}_{\mathrm{NN},1-\alpha}(\Theta^{L})&= \{\bm{\theta}\in \Theta^{L} \mid 2\big(\log(\mathcal{L}_{\mathrm{NN}}(\hat{\bm{\theta}} \mid \bm{y})) -\log(\mathcal{L}_{\mathrm{NN}}(\bm{\theta} \mid \bm{y}))\big)\leq c_{\alpha}\}\\
& = \{\bm{\theta}\in \Theta^{L} \mid 2\big(\log(p(\bm{y})\hat{\psi}(\bm{y},\hat{\bm{\theta}}))- \log(p(\bm{y})\hat{\psi}(\bm{y},\bm{\theta}))\big)\leq c_{\alpha}\}\\
& =  \{\bm{\theta}\in \Theta^{L} \mid 2\big(\log(\hat{\psi}(\bm{y}, \hat{\bm{\theta}})) - \log(\hat{\psi}(\bm{y},\bm{\theta}))\big)\leq c_{\alpha}\},
\end{aligned}
\label{eqn:gridconfidencesetnn}
\end{equation}
because the proportionality constant $p(\bm{y})$ cancels out in \eqref{eqn:gridconfidencesetnn}.

There are two reasons why we construct confidence regions. First, we would like to understand how reliable the neural likelihood surfaces are in terms of providing accurate uncertainty quantification for parameter estimation as compared to the exact or approximate likelihoods. Often, methods utilizing neural networks for statistical inference focus purely on parameter point estimation as a prediction task and generally do not provide a measure of uncertainty with their point estimates. If shown to be reliable, confidence regions constructed from the neural likelihood provide a principled method of uncertainty quantification for parameter estimation. In Section \ref{sec:casestudies}, we demonstrate the reliability of these confidence regions by analyzing their empirical coverage and size across the parameter space $\Theta$. Second, these confidence regions provide a sanity check for whether the neural likelihood is an accurate representation of the exact likelihood when the latter is unavailable. If the empirical coverage of the confidence sets derived from the neural likelihood surfaces for a sufficiently large number of realizations across the parameter space matches the expected coverage, then that increases our confidence that the neural likelihood is an accurate representation of the exact likelihood.

\section{Case Studies}
\label{sec:casestudies}

While our method of learning the likelihood function is applicable to any spatial process for which fast simulation is possible, we are most interested in providing inference for spatial processes with intractable likelihoods. Yet, in the intractable case, we can not directly compare the neural and exact likelihood surfaces. To address this issue, we first demonstrated that our method works as expected using a popular spatial process with a tractable likelihood, a Gaussian process, in our first case study. 
In our second case study, we applied our method to a spatial process with an intractable likelihood, a Brown--Resnick process, and compared the neural likelihood to the pairwise likelihood, a well-established approximation for the exact likelihood.

\subsection{First Case Study: Gaussian Process with Computationally Intensive Likelihood}
\label{sec:GPcasestudy}
The primary purpose of this case study is to validate our method in a case where the exact likelihood is available. A secondary purpose is to demonstrate the computational benefits of our method. Gaussian processes are often used to model spatial data, yet the time complexity of exact likelihood scales cubically with the number of spatial locations as described in Section~\ref{sec:backgroundspatial}. As such, developing a more computationally efficient method that provides a comparable quality of parameter estimation and uncertainty quantification as exact likelihood is an important issue in of itself. As shown in this section, our method of learning the likelihood provides such an approach.

\subsubsection{Description of Gaussian Process and Its Exact likelihood}

We applied our method to a Gaussian process with a zero mean function and an exponential covariance function with two positive parameters, length scale $\ell$ and variance $\nu$. Hence, the model~is
\begin{equation}
\begin{aligned}
f&\sim \textrm{GP}(m(\bm{x}),k(\bm{x},\bm{x'})), \textrm{ where } m(\bm{x})=0 \textrm{ and } k(\bm{x},\bm{x'})=\nu\exp(-\frac{\norm{\bm{x}-\bm{x'}}_{2}}{\ell}).
\end{aligned}
\label{eqn:gpkernel}
\end{equation}
For a realization $\bm{y}=f(\mathcal{S})$, where $\mathcal{S}$ is a regular finite grid $\mathcal{S}$ of $s$ locations with covariance matrix $\bm{\Sigma} = \big(k(\bm{x}_{i},\bm{x}_{j})_{i,j}\big)\in \mathbb{R}^{s\times s}$, the log likelihood is
\begin{equation}
\log(\mathcal{L}(\bm{\theta}\mid \bm{y})) = -\frac{1}{2}\bm{y}^\intercal \bm{\Sigma}^{-1}\bm{y} - \frac{s}{2}\log(2\pi) -\frac{1}{2}\log(\det(\bm{\Sigma})),
\label{eqn:gpll}
\end{equation}
where $\bm{\theta}=(\nu,\ell)$ \citep[pp. 81--96]{Rasmussen}.

The two most computationally intensive operations in \eqref{eqn:gpll} are computing the inverse and the determinant of the covariance matrix $\bm{\Sigma}$ each with a time complexity of $\mathcal{O}(s^{3})$. To efficiently compute both of these operations, we utilize the Cholesky decomposition of the symmetric, positive-definite matrix $\bm{\Sigma}$
\citep[pp. 202--203]{Rasmussen}. While Cholesky decomposition is computationally expensive with a time complexity of $\mathcal{O}(s^{3})$, substituting the Cholesky decomposition for the covariance matrix reduces the number of $\mathcal{O}(s^{3})$ operations from two to one. 

\subsubsection{Experiments}
\label{sec:GPexperiment}

\paragraph{\textit{Input Details}}
Since we are using a CNN as our classifier, the input shape for the spatial field and parameter must be fixed. The spatial field is a realization $\bm{y}\in \mathbb{R}^{25\times 25}$ of the spatial process for a given parameter on a $25\times 25$ regular grid $\mathcal{S}$ with spatial domain $\mathcal{D} =[-10,10]\times [-10,10]$. The parameter $\bm{\theta}$ is two-dimensional, and the evaluation parameter space $\Theta$ is $\Theta=(0,2)\times (0,2)$. The training parameter space $\tilde{\Theta}$ was extended beyond the evaluation parameter space $\Theta$ to $\tilde{\Theta}=(0, 2.5) \times (0, 2.5)$ to better learn the likelihood and prevent potential biases at the boundary. For more training details, see Section \ref{supplement:traingp} of the Supplementary Material.

\paragraph{\textit{Simulating Training and Validation Data}} To generate the training data, we simulated $n=500$ spatial field realizations for each of the $m=3000$ parameters sampled from the training parameter space $\tilde{\Theta}$ via Latin hypercube sampling as explained in Section \ref{sec:methodclasses}. Separately from the training data, we simulated $n=500$ spatial field realizations for each of m=300 parameters sampled from $\tilde{\Theta}$ via Latin hypercube sampling to serve as the validation data.

\paragraph{\textit{Architecture}}
Here we provide details about the CNN architecture described in Section \ref{sec:methodsarchitecture}. An ideal architecture is an architecture flexible enough to learn the likelihood for a large variety of spatial processes. To achieve such versatility, the CNN architecture must be complex enough to be able to learn a potentially rough, non-convex likelihood yet not too complex to overfit when learning a smooth, convex likelihood. While the ability to learn the likelihood for any spatial process is most likely too ambitious, the architecture we propose does have the flexibility to learn the likelihoods for both Gaussian and Brown--Resnick processes. This architecture is motivated by the architecture of the CNN trained for neural estimation in \citet{Lenzi}.

In the first part of the architecture, there are three convolutional layers with a decreasing number of filters starting with 128 filters for the first, 128 filters for the second, and 16 for the third. All the convolutional layers have the same kernel size and activation function which are $3\times 3$ and ReLU, respectively. Between these convolutional layers are max pooling layers with the same kernel size of $2\times 2$. The convolutional part outputs a flatten vector of size $64$ to which we concatenate the parameter $\bm{\theta}\in \mathbb{R}^{2}$. In the fully connected part of the architecture, there are 4 layers with a decreasing number of neurons and the same ReLU activation function except for the last layer which has a softmax activation function to handle probabilistic binary classification. The output is a 2-d vector in which the first entry is the predicted probability $\hat{h}(\bm{y},\bm{\theta}) = \hat{P}(C=1 \mid (\bm{y}, \bm{\theta}))$. See Table~\ref{Tab:architecture} in the Supplementary Materials for the specific architecture.

\paragraph{\textit{Training and Model Selection}} We trained several models with varying configurations of batch size and learning rate schedule for twenty epochs. We selected a model with the lowest validation loss amongst the versions in which there was little to no indication of overfitting in the plot of training and validation loss. The selected model was trained with a batch size of $30000$ and a learning rate schedule in which the the learning rate started at $0.001$ for the first five epochs and decreased by a multiplicative factor of $e^{-0.1}$ for each epoch after the first five. Due to the large batch size, we distributed training over four NVIDIA Tesla K80 GPUs at the most to address memory requirements for such a large matrix (i.e., $30000\times 25 \times 25)$.

\paragraph{\textit{Calibration}}

First, we generated training data by simulating $n=50$ realizations per each of the $m=3000$ parameters sampled via Latin hypercube sampling from the evaluation parameter space $\Theta$.
 With the same process, we generated test data: $n=50$ realizations per each of the $m=300$ sampled parameters. The training and test data for calibration were independent of the data used to train the CNN. After constructing the two classes for both training and test data, we used the uncalibrated CNN to obtain the corresponding classifier outputs. These classifier outputs coupled with the corresponding class labels formed the training and test data for the logistic regression model in \eqref{eqn:plattscale}.
 
 To determine the effectiveness of calibration, we examined reliability diagrams as well as the neural likelihood surfaces before and after calibration for the test data. The reliability diagram depicts the empirical class probability as a function of the predicted class probability \citep{Guo}. For a perfectly calibrated model, the reliability diagram should show the identity function. Thus, the closer the calibrated probabilities are to the identity function, the more effective the calibration.

\begin{figure}[!t]
    \centering  \includegraphics[scale = .3]{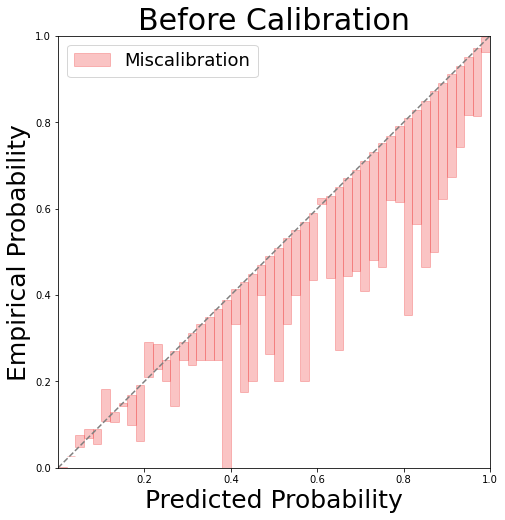}
    \includegraphics[scale = .3]{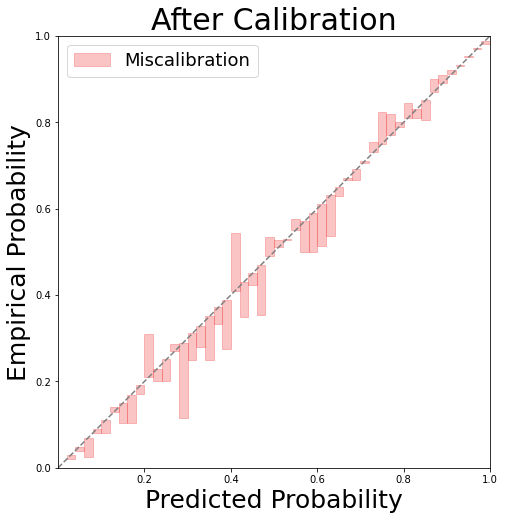}
    \caption{Reliability diagram for class one before calibration (left) and after calibration (right) for Gaussian process.}
    \label{fig:gpreliability}
\end{figure}

The reliability diagram after calibration in Figure~\ref{fig:gpreliability} shows great improvement in how closely the predicted and true class probabilities match. Before calibration, the empirical class one probability is generally lower than expected for any given predicted class one probability. Calibration fixes this tendency
for the classifier to overestimate the probability a given realization and parameter is in the first class, the class in which the parameter $\bm{\theta}$ and realization $\bm{y}$ are dependent. Since learning the likelihood relies on estimating this probability correctly, fixing this tendency produces calibrated neural likelihood surfaces in which the area of high likelihood is larger and improves approximate confidence regions and empirical coverage as we demonstrate later in this section.

\paragraph{\textit{Evaluation}}
 We discuss now in detail how we evaluated the neural likelihood's performance as compared to standard methods. First, we created an evaluation dataset by generating  $n = 200$ spatial field realizations per parameter on a $9\times 9$ regular grid over the evaluation parameter space $\Theta = (0,2)\times (0,2)$ for both spatial processes. If there are certain regions in which the CNN underperforms, we can detect these regions when evaluating surfaces, parameter estimators, approximate confidence regions, and empirical coverage with this evaluation data.
 

\paragraph{\textit{Evaluation: Surfaces}}
Using the process described in Section \ref{sec:methodinferencesurface}, we constructed the neural and exact likelihood surfaces with a $40\times 40$ regular grid $\Theta^{L}$ in \eqref{eqn:surfacegrid} over the evaluation parameter space $\Theta = (0,2)\times (0,2)$. As mentioned in Section~\ref{sec:methodinferencesurface}, the surfaces are truly the log likelihood surfaces which ensures the ranges of the exact and neural likelihoods differ by an additive constant rather than a multiplicative constant. For visualizations, we utilized a color scale in which the maximum values of both the color scale
and the particular surface match to deal with the shifted range due to the additive constant. We selected a range of ten units for the color scale because an approximate confidence region in a two-dimensional parameter space has $99\%$ coverage probability for a cut-off value $C_{.01}=9.21$, the $99\%$ quantile of a $\chi^2$ distribution with two degrees of freedom in \eqref{eqn:confidenceset}.

\begin{figure}[!t]
    \centering
    \includegraphics[scale = .17]{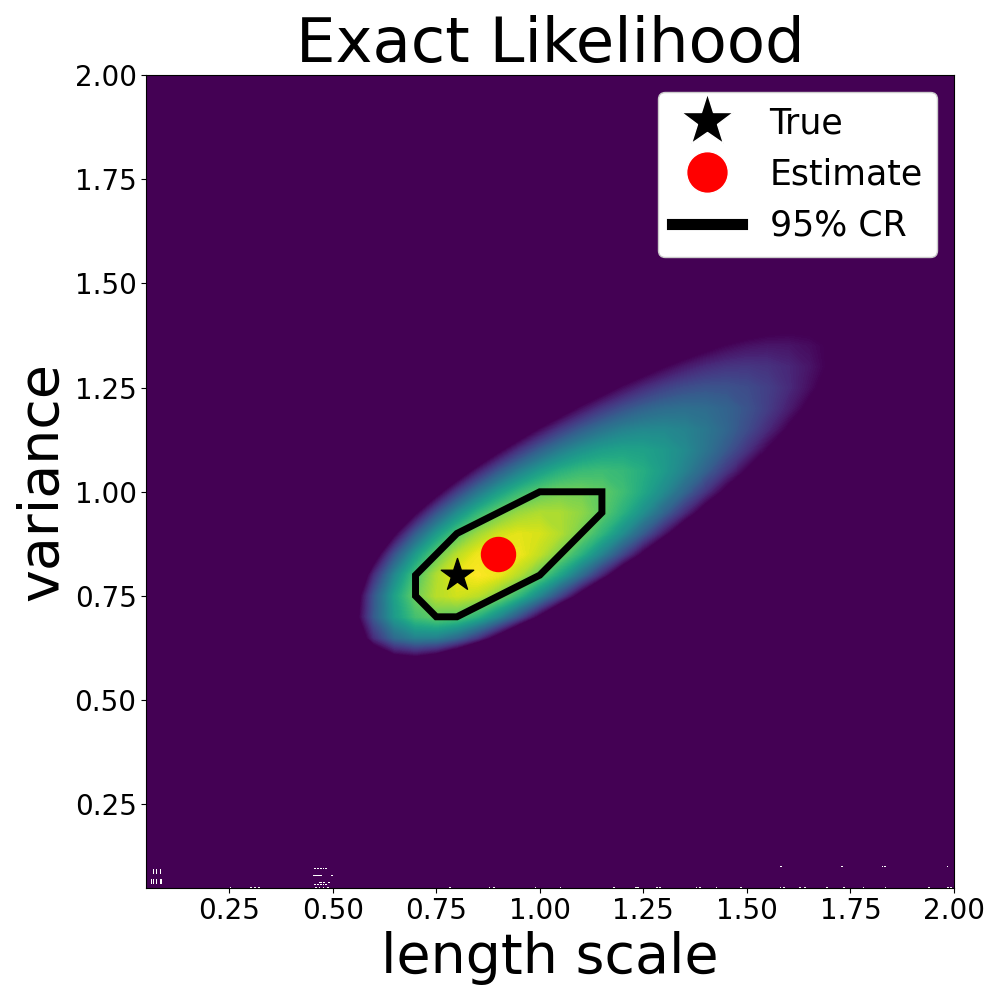}
    \includegraphics[scale = .17]{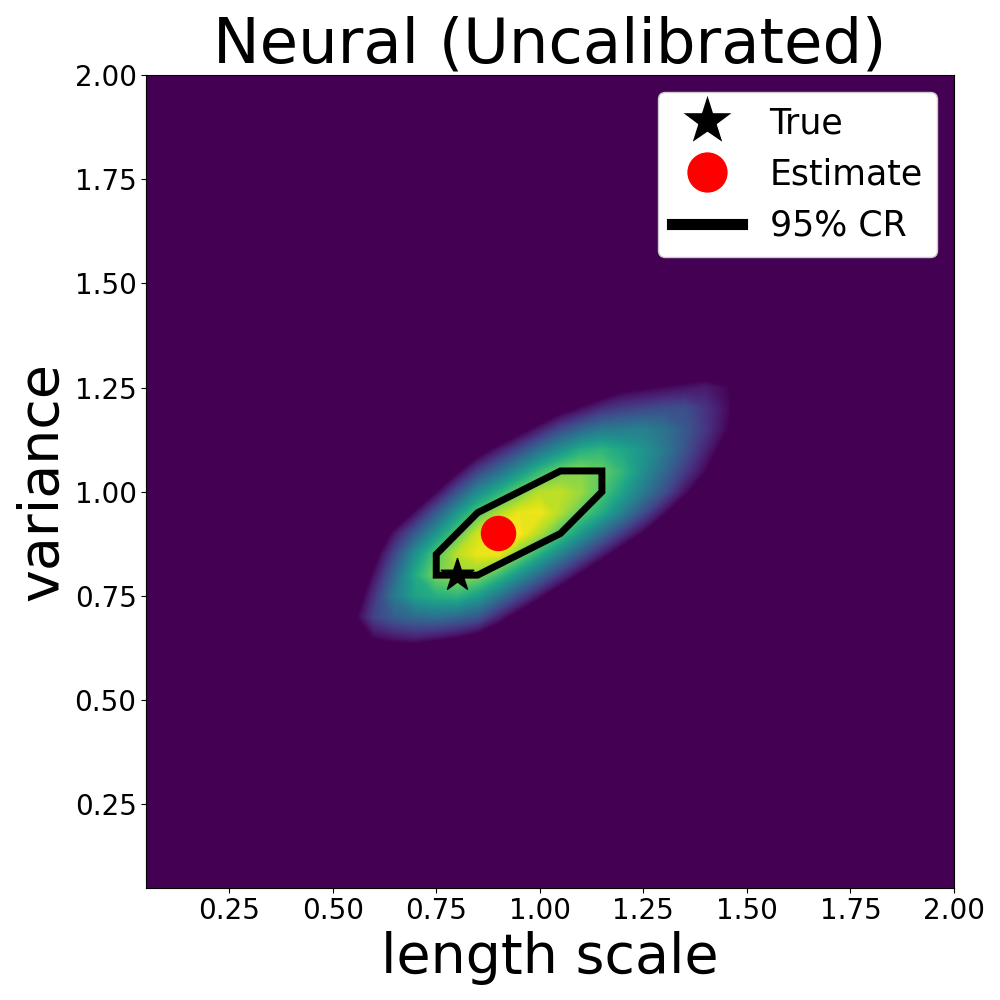}
    \includegraphics[scale = .17]{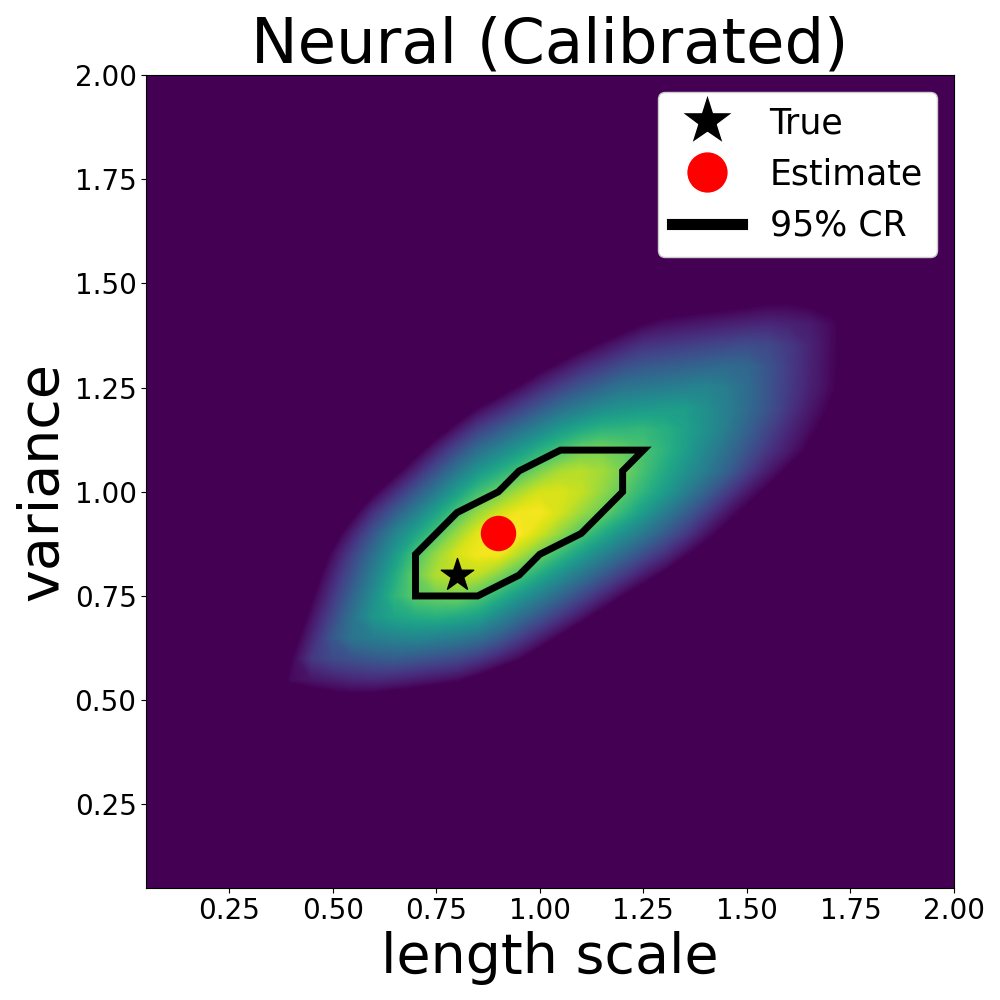}
    \caption{Surfaces for exact log exact likelihood (left), neural likelihood before calibration (center), and neural likelihood after calibration (right) for a realization of a Gaussian process with parameters $\nu = 0.8$ and $\lambda = 0.8$. In each figure, the color scale ranges from the maximum value of the surface to ten units less than the maximum value.}
    \label{fig:gpsurfaces}
\end{figure}

Across the evaluation parameter space $\Theta$, the areas of high likelihood for the neural and exact likelihood surfaces are similar in shape and location yet differ in size. In correcting the classifier's tendency to overestimate the class one predicted probability, calibration spreads out the high-likelihood region. Consequentially, the neural likelihood surfaces more closely resemble the exact likelihood surfaces in terms of the size of the high-likelihood region. Yet, the area of high likelihood is slightly too large after calibration. See Figure~\ref{fig:gpsurfaces} for a visualization of these observations which hold true for all surfaces of spatial field realizations generated for any parameter in~$\Theta$. 

\paragraph{\textit{Evaluation: Parameter Estimation}}

\begin{figure}[!t]
    \centering
    \includegraphics[scale = .30]{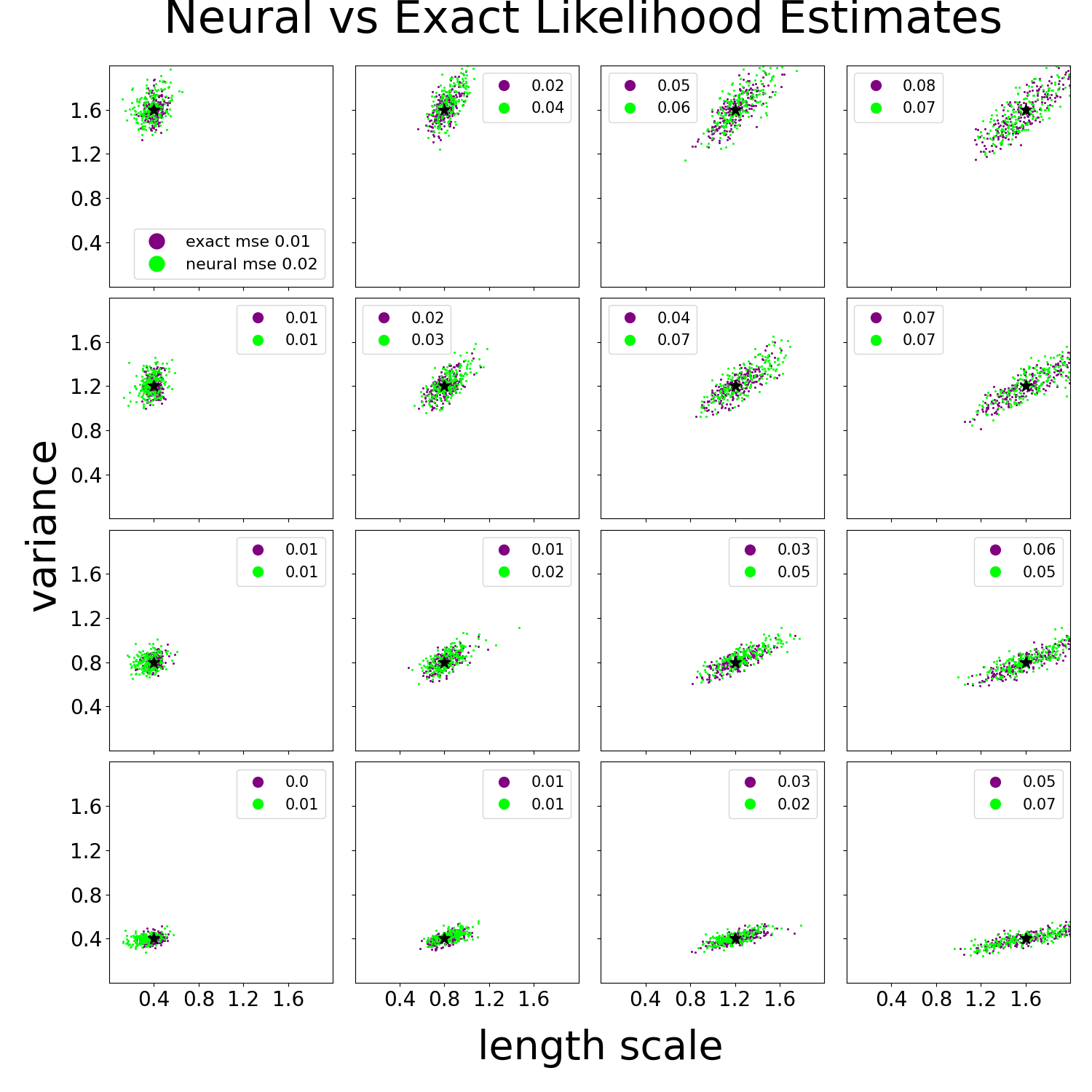}
    \caption{Parameter estimates for exact and neural likelihood for a Gaussian process. Each of the $16$ plots contains the true parameter (black star) which generated the $200$ spatial field realizations and the corresponding parameter estimates for exact likelihood (purple) and neural likelihood (green) with mean squared error (MSE) in the legend. The true parameter increases in variance from bottom to top and in length scale from left to right.}
    \label{fig:gpparams} 
\end{figure}

\begin{figure}[!t]
    \centering
    \includegraphics[scale = .30]{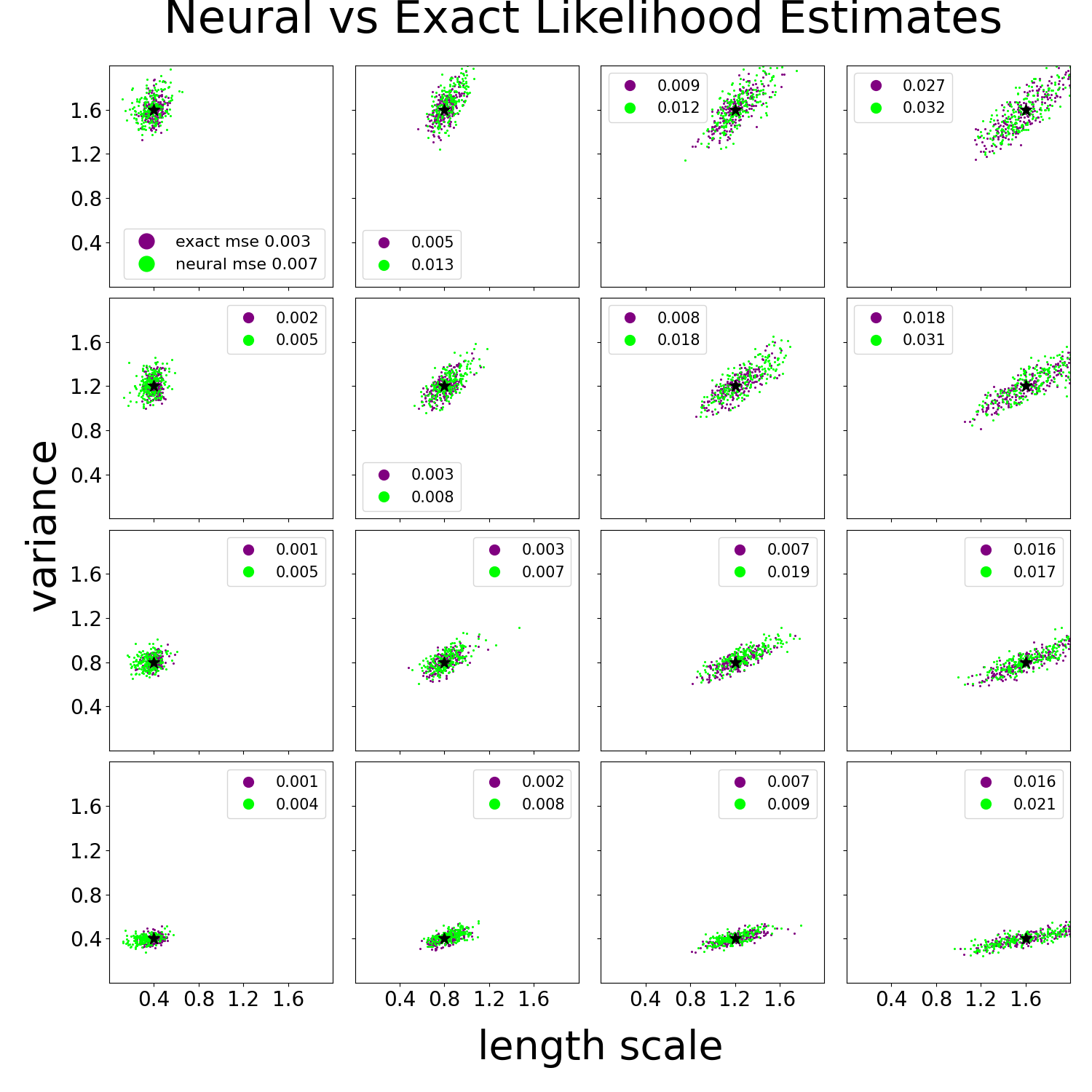}
    \caption{Parameter estimates for exact and neural likelihood in the case of 5 i.i.d.\ spatial field realizations for a Gaussian process. Each of the $16$ plots contains the true parameter (black star) which generated the realizations and the corresponding parameter estimates for exact likelihood (purple) and neural likelihood (green) with mean squared error (MSE) in the legend. The true parameter increases in variance from bottom to top and in length scale from left to right.}
    \label{fig:gpfiveparams}
\end{figure}

To compare the parameter estimates for neural and exact likelihood, we display a $4\times 4$ grid of plots in which each plot contains the parameter estimates from both methods for $200$ realizations generated from a single parameter in the evaluation data. The parameter estimates for the neural likelihood before and after calibration are the same because Platt scaling is a monotonic transformation. Since the parameter estimates are all on the same $40\times 40$ grid, the estimates are jittered with a small amount of noise in Figures~\ref{fig:gpparams} and \ref{fig:gpfiveparams} to distinguish the individual estimates. The parameters which generated the realizations range from $0.4$ to $1.6$ by increments of $0.4$ for both parameters.


From Figure~\ref{fig:gpparams}, we conclude that the parameter estimates for both methods are comparable across the evaluation parameter space $\Theta$. This indicates that the neural likelihood serves as a great surrogate for the exact likelihood in terms of point estimation. The neural likelihood estimator $\hat{\bm{\theta}}_{NN}$ has greater variance than the exact likelihood estimator $\hat{\bm{\theta}}$, yet the patterns of behavior are the same for both estimators across $\Theta$. For instance, as the true length scale $\ell$ increases, both exact and neural likelihood estimators increase in variance along the length scale axis. 

\paragraph{\textit{Evaluation: Approximate Confidence Regions}}

We constructed $95\%$ approximate confidence regions for each realization in the evaluation data. As shown in Figure~\ref{fig:gpsurfaces}, the neural likelihood approximate confidence regions become larger yet retain their shape and location after calibration because Platt scaling is a monotonic transformation. 
After calibration, the neural and exact likelihood confidence regions are more comparable in shape, location, and size because the neural likelihood confidence regions increase in size from $60\%$ to $120\%$ of the size of the exact likelihood confidence regions on average.

\paragraph{\textit{Evaluation: Empirical Coverage and Confidence Region Area}}
We computed empirical coverage and confidence region area for the evaluation data. For each of the two hundred realizations $\{\bm{y}_{j,l}\}_{j\in [200]}$ per parameter $\bm{\theta}_{l}$ on the $9\times 9$ grid, we determined whether the true parameter $\bm{\theta}_{l}$ is in the $95\%$ approximate confidence region as well as the area of the region. The empirical coverage for the parameter $\bm{\theta}_{l}$ is \begin{equation}
    \frac{1}{n}\sum_{j=1}^{n} \mathbf{1}\big(\bm{\theta}_{l}\in \mathcal{C}_{\textrm{NN},.95}(\Theta^{L},\bm{y}_{j,l})\big),
\end{equation} where $n=200$. We visualize the empirical coverage and confidence region area with a heat map over the evaluation parameter space $\Theta$ to determine whether the coverage and area vary over $\Theta$.

\begin{figure}[!t]
    \centering
    \includegraphics[scale = .22]{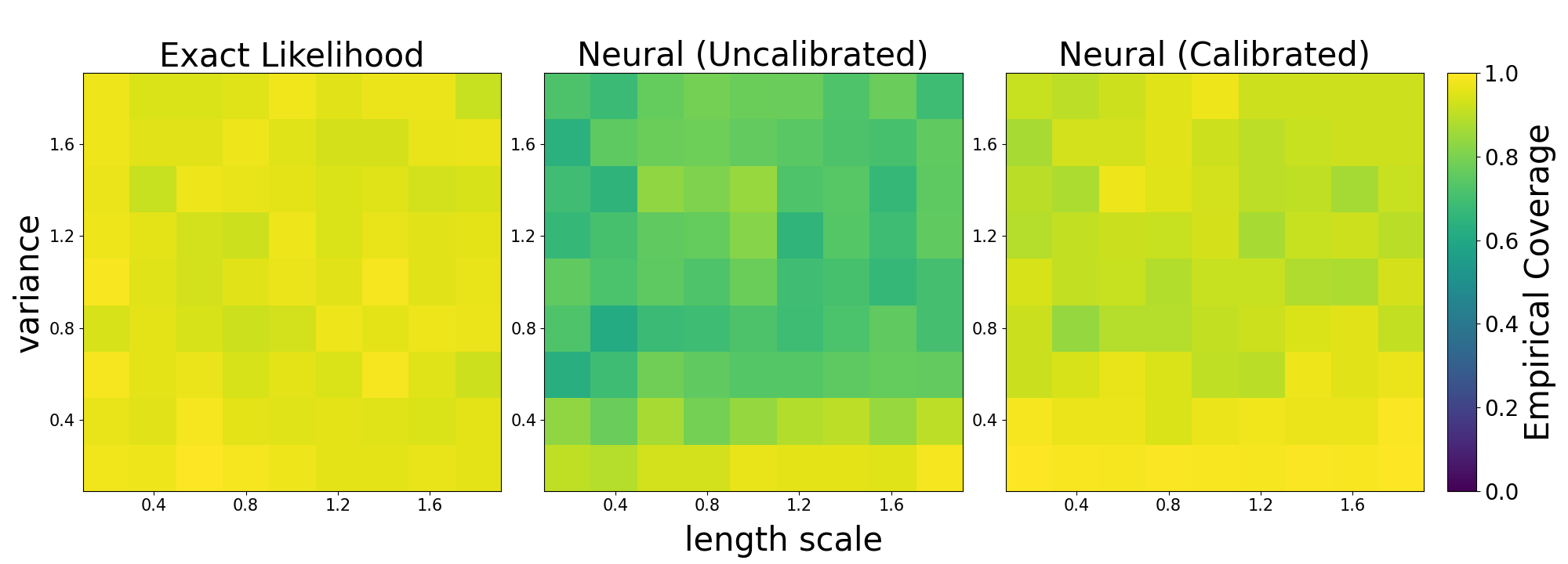}
    \caption{Empirical coverage for $95\%$ approximate confidence regions for the exact likelihood (left) and neural likelihood before calibration (center) and after calibration (right) for $200$ realizations per parameter on a $9\times 9$ grid over $\Theta$.}
    \label{fig:gpempiricalcoverage}
\end{figure}

\begin{figure}[!t]
    \centering
    \includegraphics[scale = .22]{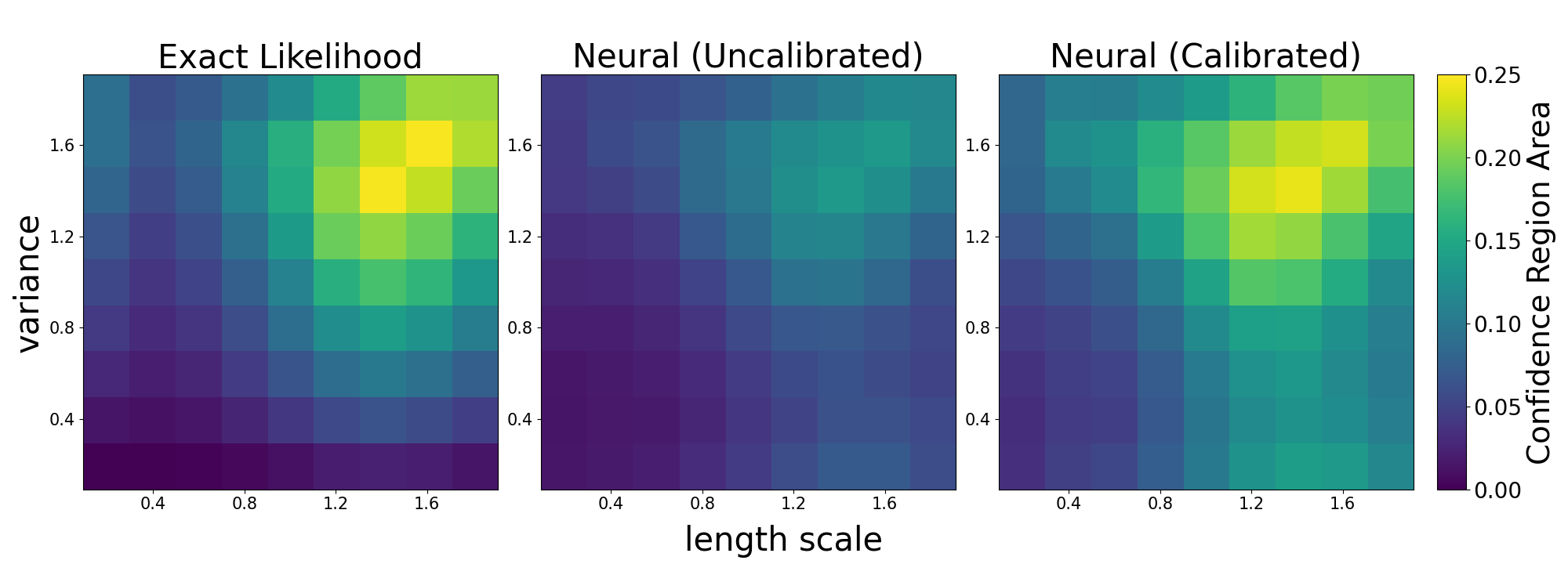}
    \caption{$95\%$ approximate confidence region area for exact likelihood (left) and neural likelihood before calibration (center) and after calibration (right) for $200$ realizations per parameter on a $9\times 9$ grid over $\Theta$.}
    \label{fig:gpcrarea}
\end{figure}

As shown in Figure~\ref{fig:gpcrarea}, calibration increases average confidence region area for neural likelihood from $60\%$ to $120\%$ of the average area for exact likelihood. Confidence region area for exact and neural likelihood become more comparable after calibration, and thus, empirical coverage for neural likelihood increases from $80\%$ to approximately $93\%$ across $\Theta$ as shown in  Figure \ref{fig:gpempiricalcoverage}. Additionally, the coverage for both exact and neural is essentially uniform across $\Theta$. 

\paragraph{\textit{Evaluation: Timing Study}}

We conducted a timing study for evaluating the neural and exact likelihood surfaces using an Intel Core i7-10875H processor with eight cores, each with two threads. For fifty realizations $\bm{y}_{i}\in \mathbb{R}^{25\times 25}$, we recorded the average elapsed time and standard deviation in evaluating the neural or exact likelihood surface on a $40\times 40$ grid over the evaluation parameter space $\Theta$. As mentioned in Section \ref{sec:methodinferencesurface}, the CNN can handle vectorized inputs which significantly accelerates evaluation of the neural likelihood surface. To understand the impact of vectorization, we timed the evaluation of the neural likelihood when using both vectorized and unvectorized inputs.

As far as we are aware, there is no prepackaged way to vectorize the evaluation of the exact likelihood. As such, when conducting a similar study for the exact  likelihood, we used Cholesky factorization and multicore processing but no vectorization when evaluating the exact likelihood for the same fifty realizations using the same processor.

As shown in Table~\ref{Tab:gptime}, vectorizing reduces the time to compute the neural likelihood surface by a factor of approximately 6. Due to a discrepancy in the ability to handle vectorized computations, the time difference between constructing neural and exact likelihood surfaces is significant. The vectorized neural likelihood surfaces are approximately thirty times faster to produce than exact likelihood surfaces. Since the computational cost of exact likelihood increases cubically with the number of spatial locations, we expect that the computational efficiency of vectorized neural likelihood will only increase in comparison to exact likelihood as the number of locations increases.

We did not include the upfront cost of neural network training in the recorded times of evaluating the neural likelihood surface as we are considering the ideal use case for this method---evaluating the amortized neural likelihood repeatedly under the same inference framework in which case the upfront cost of training diminishes in importance. 
However, for reference, training the neural network on a cluster of 4 NVIDIA Tesla K80 GPUs with the given data and batch sizes took less than ten hours.


\begin{table}[!t]
\caption{Time to produce neural and exact likelihood surfaces on a $40\times 40$ grid over $\Theta$ for $50$ realizations of a Gaussian process on a $25\times 25$ grid with a spatial domain of $[-10,10]\times [-10,10]$.}
\small
\begin{center}
\begin{tabular}{ |p{6cm} p{2.5cm} p{3cm}| }
 \hline
 
 Type of surface and method & average (sec) & std. dev. (sec)\\
 \hline
 exact likelihood & 71.67 & 1.02\\
  unvectorized neural likelihood & 13.46 & 0.26\\
 vectorized neural likelihood  & 2.26 &  0.34\\
 \hline
\end{tabular}
\end{center}
\label{Tab:gptime}
\end{table}

\paragraph{\textit{Evaluation: Multiple Realizations}}

As explained in Section \ref{sec:methodslikelihoodconnection}, our method of learning the likelihood is applicable to the case of an arbitrary number of $\textrm{i.i.d.}\ $realizations $\bm{y}_{1}
\hdots \bm{y}_{n}$ from the spatial process without needing to retrain the single-realization neural network. To demonstrate this, we generated evaluation data for five $\textrm{i.i.d.}\ $spatial field realizations in the same manner as we did in the single realization case and evaluated neural likelihood according to \eqref{eqn:multiclassificationconnection} for the evaluation data. We display the $4\times 4$ plot of the resulting parameter estimates  as we do for the single realization~case.

The parameter estimators for exact and neural likelihood in the multiple i.i.d.\ realizations case in Figure~\ref{fig:gpfiveparams} are more accurate and have less variance than equivalent estimators for the single realization case in Figure~\ref{fig:gpparams}, as expected. As in the single realization case, the neural likelihood estimator $\hat{\theta}_{NN}$ has slightly greater variance than the exact likelihood estimator $\hat{\theta}$, yet the patterns of behavior are similar for both estimators across the evaluation parameter space $\Theta$.

\paragraph{\textit{Summary of Results}} From this case study, we demonstrated that our method of learning the likelihood is both accurate and computationally efficient for Gaussian processes. Once calibrated, the neural likelihood surfaces, parameter estimates, and approximate confidence regions are comparable to the equivalent for exact likelihood. From the evaluation of neural likelihood surfaces, confidence regions, and empirical coverage before and after calibration, it is clear that calibration is essential to our method in order to achieve results comparable to exact likelihood. Finally, the vectorized neural likelihood surfaces are significantly faster to evaluate than exact likelihood surfaces because neural networks are fast to evaluate and our CNN, once trained, is amortized. Thus, we have both validated our method in a case where the exact likelihood is available and shown this method is more computationally efficient than exact likelihood.

\subsection{Second Case Study: Brown--Resnick Process with an Intractable Likelihood}
\label{sec:casestudytwo}
We next apply our method to a Brown--Resnick process which has an intractable likelihood. Yet, there is composite likelihood, an approximation for the exact likelihood \citep{Castruccio}, to which we can compare neural likelihood. 
Our selection of a Brown--Resnick Process for this case study is in part due to its use as a case study for neural estimation in \citet{Lenzi}. 

\subsubsection{Description of Brown-Resnick Process and its Approximate Likelihood}
\label{sec:brdescription}

The Brown--Resnick process has two parameters--$\lambda\in \mathbb{R}^{+}$, a range parameter which determines the degree to which the spatial process at locations a fixed distance away impact the spatial process at a given location, and $\nu \in (0,2]$, a smoothness parameter which determines the overall degree of smoothness of the process across the spatial domain. As such, the parameter space $\Theta$ for which we are interested in learning the likelihood is a bounded subset of $\mathbb{R}^{+}\times (0,2]$.

The likelihood involves a summation over the set of all possible partitions of the spatial locations $\{\bm{s}_{i}\}_{i\in [n]}$ at which the Brown-Resnick process is observed \citep{Castruccio}. This set has cardinality equal to the bell number $B_{n}$ which grows more than exponentially with respect to $n$, the number of spatial locations. Since the number of spatial locations is generally large in practice, computing the likelihood for a Brown--Resnick process is intractable in many practical cases.

Yet, in the bivariate case of only two spatial locations, the likelihood has a simple closed form. Using the bivariate case, we can provide an approximation for the full likelihood of a Brown--Resnick process called pairwise likelihood, a form of composite likelihood. In pairwise log likelihood, the summands are the bivariate log likelihoods between two spatial locations $\bm{s}_{j_{1}}$ and $\bm{s}_{j_{2}}$ for $j_{1},j_{2}\in [n]$, and involve only select pairs $(\bm{s}_{j_{1}},  \bm{s}_{j_{2}})$ because even a summation over all $\frac{n(n-1)}{2}$ pairs is computationally intensive for a large number of locations.

Since nearby locations contain the most information about a given location, the selection criterion generally is determined by the distance between locations. When we compute pairwise likelihood in Section~\ref{sec:brexperiment}, all pairs of observations whose locations are within a certain cut-off distance $\delta$ of each other are included. The cut-off distance $\delta$ is a tuning parameter that must be appropriately selected in order to obtain reasonable results. In practice, if the cut-off distance is too small or too large, the maximum pairwise likelihood estimates can be highly inaccurate, and the pairwise likelihood surfaces tend to be uninformative as shown in Section \ref{supplement:additionalresults} of the Supplementary Material. In contrast, neural likelihood does not involve such tuning parameters.

Additionally, the asymptotic behavior of the maximum pairwise likelihood estimator and, subsequently, the pairwise likelihood ratio statistic is different than the equivalent for the full likelihood. To enable asymptotic inference, \citet{Chandler_2007} proposed adjusting the pairwise likelihood such that the asymptotic distribution of the adjusted pairwise likelihood ratio statistic is the same as that of the full likelihood. 
See Sections \ref{supplement:brdescription} and \ref{supplement:adjustment} of the Supplementary Material for further technical details on adjusting the pairwise likelihood.

\subsubsection{Experiments}
Unless otherwise stated, the details of this case study are the same as the Gaussian process case study in Section~\ref{sec:GPcasestudy}.
\label{sec:brexperiment}
\paragraph{\textit{Training and Model Selection}}

As in the Gaussian process case, the evaluation parameter space is $\Theta = (0,2)\times (0,2)$. For the Brown--Resnick process, the training parameter space $\tilde{\Theta}$ is the same as evaluation because training difficulties arise if the training parameter space is extended. The selected model was trained with a batch size of $50$ and a learning rate schedule in which the learning rate started at $0.002$ for the first five epochs and decreased by a multiplicative factor of $e^{-0.1}$ for each of the fifteen epochs after the first five. See Section \ref{supplement:trainbr} of the Supplementary Material for more detail as to why the given batch size and learning rate were selected and what training issues we encountered when extending the training parameter space.

\begin{figure}[t]
    \centering
    \includegraphics[scale = .3]{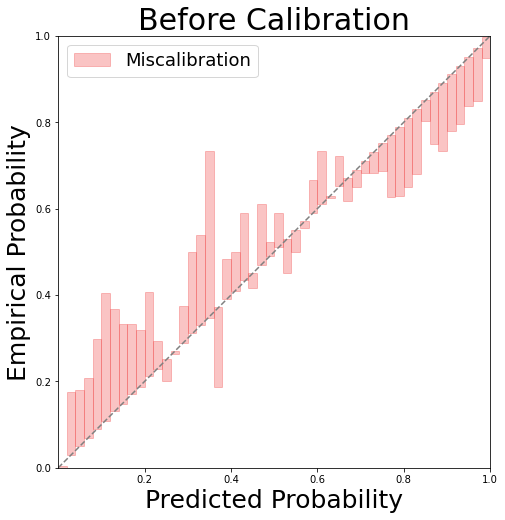}
    \includegraphics[scale = .3]{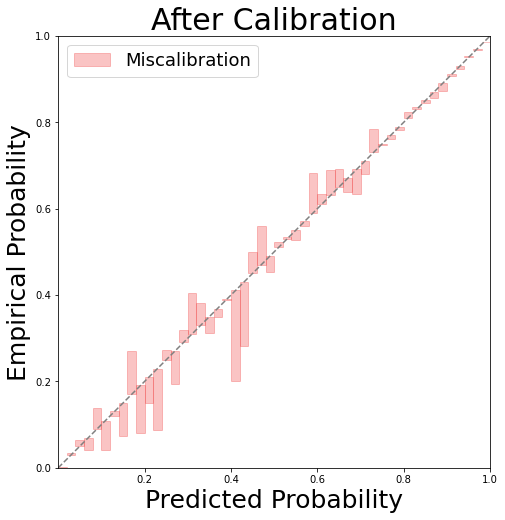}
    \caption{Reliability diagrams for class one before calibration (left) and after calibration (right) for Brown--Resnick process.}
\label{fig:brreliablity}
\end{figure}

\paragraph{\textit{Calibration}}
The reliability diagrams before and after calibration in Figure~\ref{fig:brreliablity} show great improvement in how closely the predicted and true class probabilities match after calibration. Before calibration, the classifier either underestimates or overestimates the probability of a given parameter and spatial field pair belonging to class one depending on the predicted class one probability. Calibration largely fixes these tendencies, and the resulting calibrated neural likelihood surfaces and approximate confidence regions are more accurate as shown later in this section.

\paragraph{\textit{Pairwise Likelihood}} We compare neural likelihood to both unadjusted and adjusted pairwise likelihood for distance cut-off $\delta = 2$. See Section \ref{supplement:additionalresults} of the Supplementary Material for comparisons of neural and pairwise likelihood for various distance cut-offs $\delta$. To compute pairwise likelihood, we used the \texttt{fitmaxstab} function in the R package SpatialExtremes \citep{SpatialExtremes}. The \texttt{fitmaxstab} function fits data to a max-stable process such as a Brown--Resnick process and returns a list of objects including the function \texttt{nllh}, the negative log pairwise likelihood function for the given data and weighting scheme. For each realization $\bm{y}$ in the evaluation data, we produced a pairwise likelihood surface, parameter point estimate and approximate confidence region by evaluating the \texttt{nllh} function at each parameter on the $40\times 40$ grid over $\Theta$.

\paragraph{\textit{Adjusted Pairwise Likelihood}} We adjusted the pairwise likelihood surfaces for $\delta = 2$ according to the process described in Sections \ref{supplement:brdescription} and \ref{supplement:adjustment} of the Supplementary Material. 
Adjusting the pairwise surfaces for $\delta=1$ case failed. See Section \ref{supplement:adjustment} of the Supplementary Material for details as to why the adjustment failed in this case. We do not display separate parameter estimates for the unadjusted and adjusted pairwise likelihoods because the pairwise likelihood parameter estimates are unchanged after the adjustment. As described in Section \ref{supplement:adjustment} of the Supplementary Material, there are two methods available to perform the adjustment---Cholesky factorization and eigendecomposition. In this paper, we display only results using Cholesky factorization as this method produced the better results between the two adjustments in terms of empirical coverage and confidence region area.

\begin{figure}[!t]
    \centering
    \includegraphics[scale = .13]{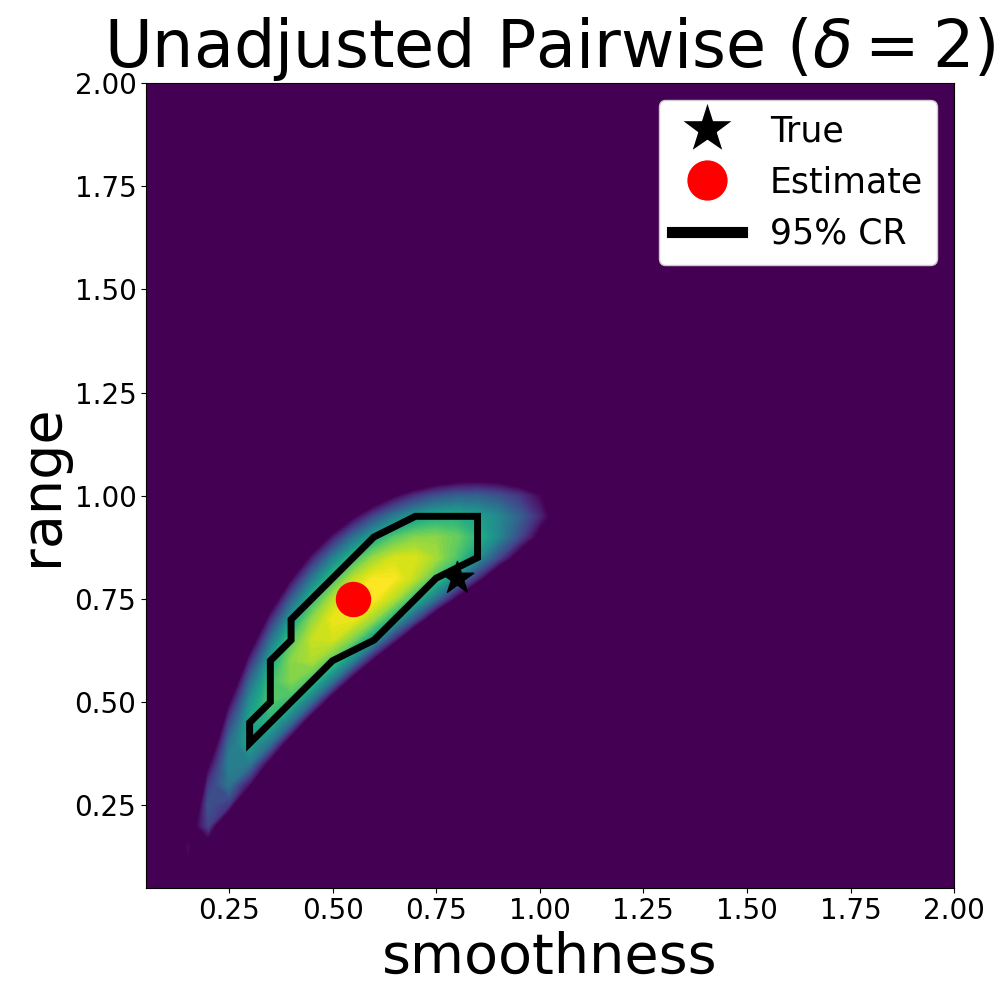}
    \includegraphics[scale = .13]{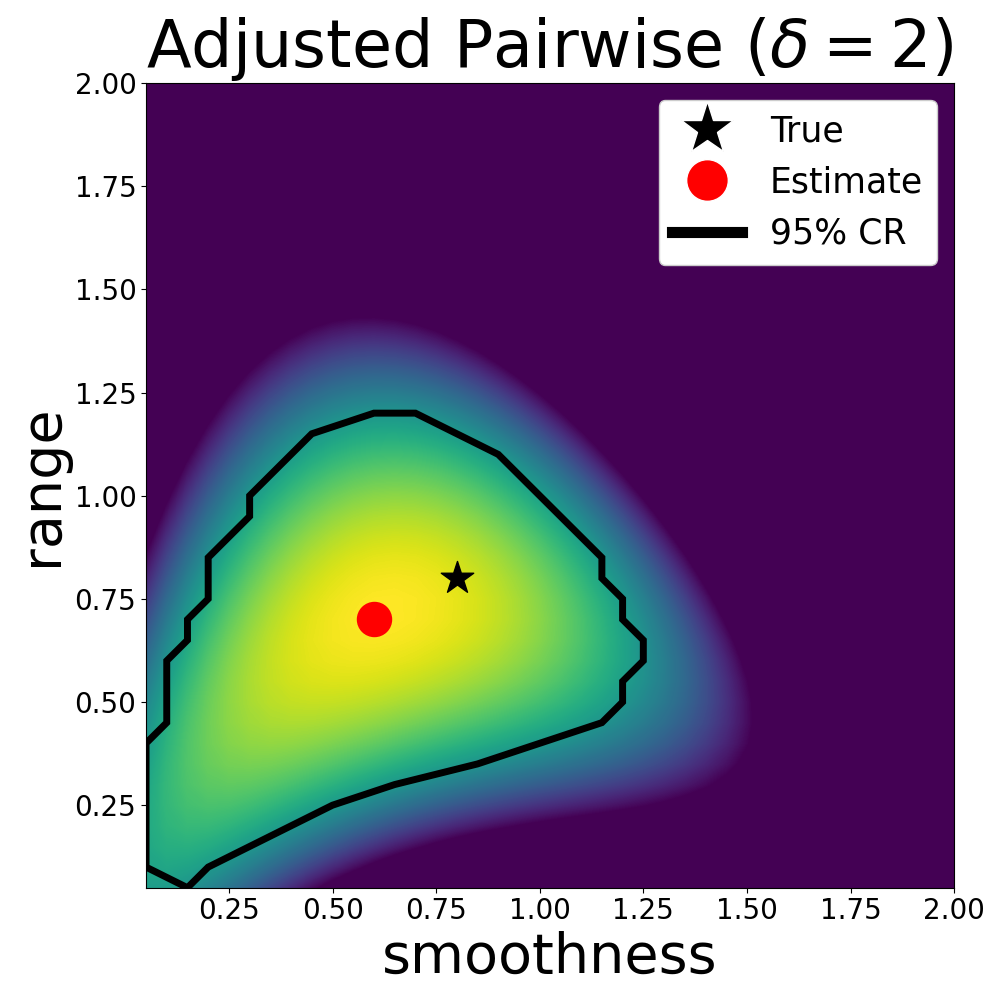}
    \includegraphics[scale = .13]{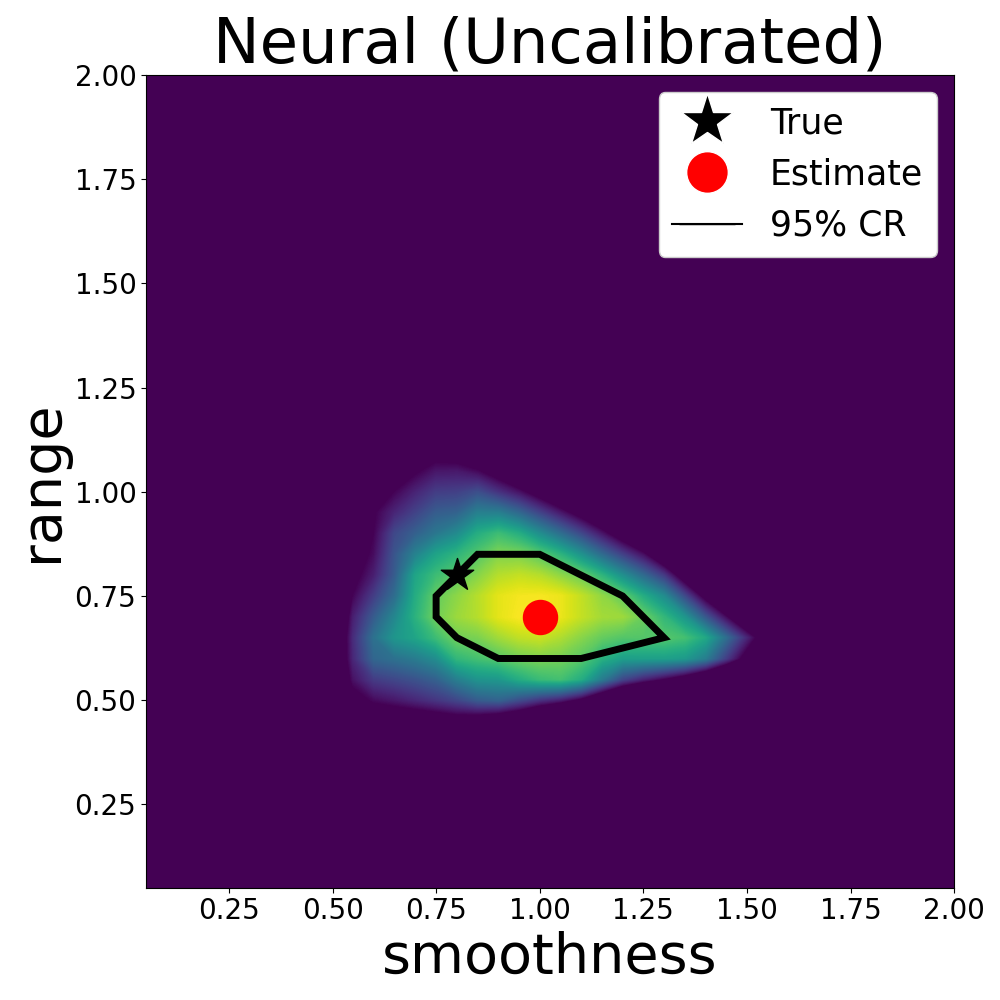}
    \includegraphics[scale = .13]{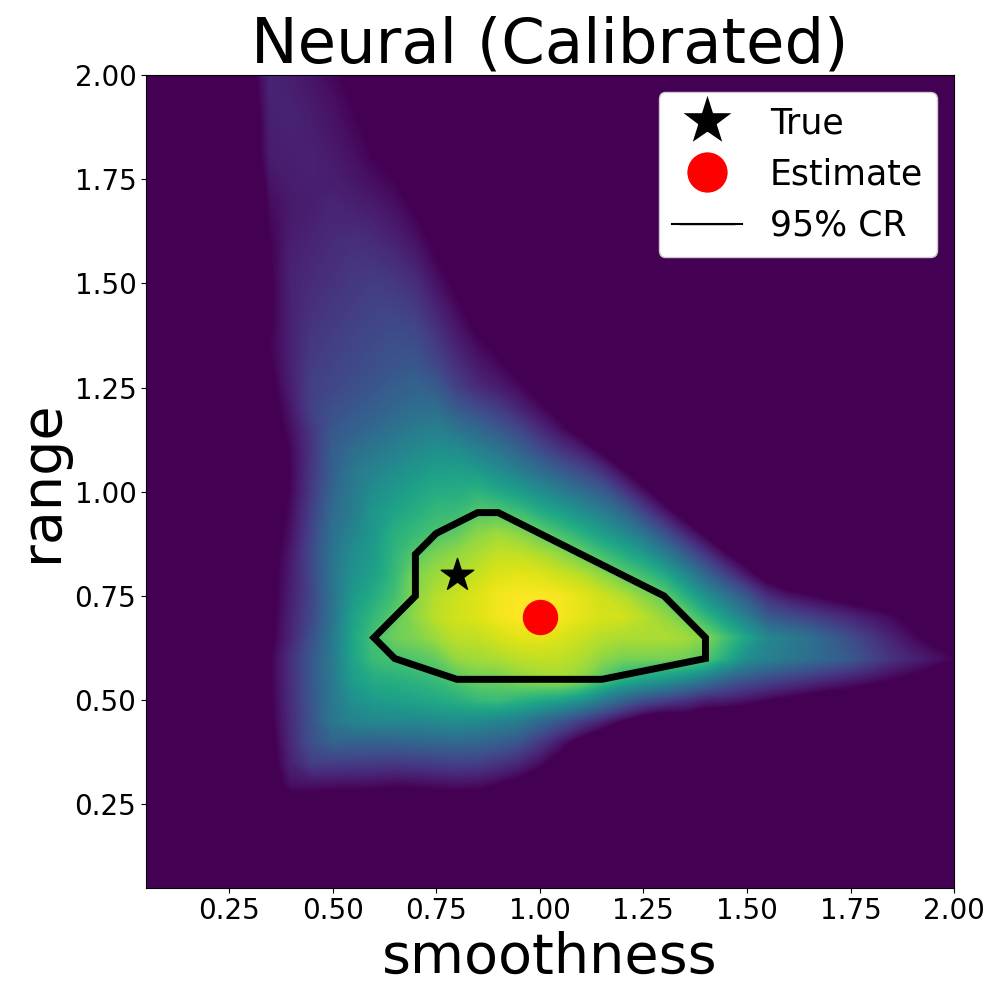}
\caption[short]{The pairwise likelihood surface for distance cut-off $\delta=2$ before adjustment (far left) and after adjustment (center left) and neural likelihood surface before calibration (center right) and after calibration (far right) for a realization of a Brown--Resnick process with parameters $\nu = 0.8$ and $\lambda = 0.8$. In each figure, the color scale ranges from the maximum value of the surface to ten units less than the maximum value.}
\label{fig:brsurfaces}
\end{figure}

\paragraph{\textit{Evaluation: Surfaces}} We compare the neural likelihood surface before and after calibration to both the unadjusted and adjusted pairwise likelihood surfaces for distance cut-off $\delta = 2$ for the same realization $\bm{y}$ in the evaluation data in Figure~\ref{fig:brsurfaces}. We do not expect the neural and pairwise likelihood surfaces, whether adjusted or not, to exactly mirror each other because pairwise likelihood is an approximation of the exact likelihood. 
In general, the area of high likelihood in the pairwise surface increases after adjustment yet, as expected, the point estimate is unchanged. The neural and pairwise likelihood surfaces, whether adjusted or not, exhibit very different behavior in shape, location, and size of the area of high likelihood. Calibration increases the area but maintains the shape and location of the high likelihood region in the neural likelihood surface.

\begin{figure}[!t]
    \centering
        \includegraphics[scale = .30]{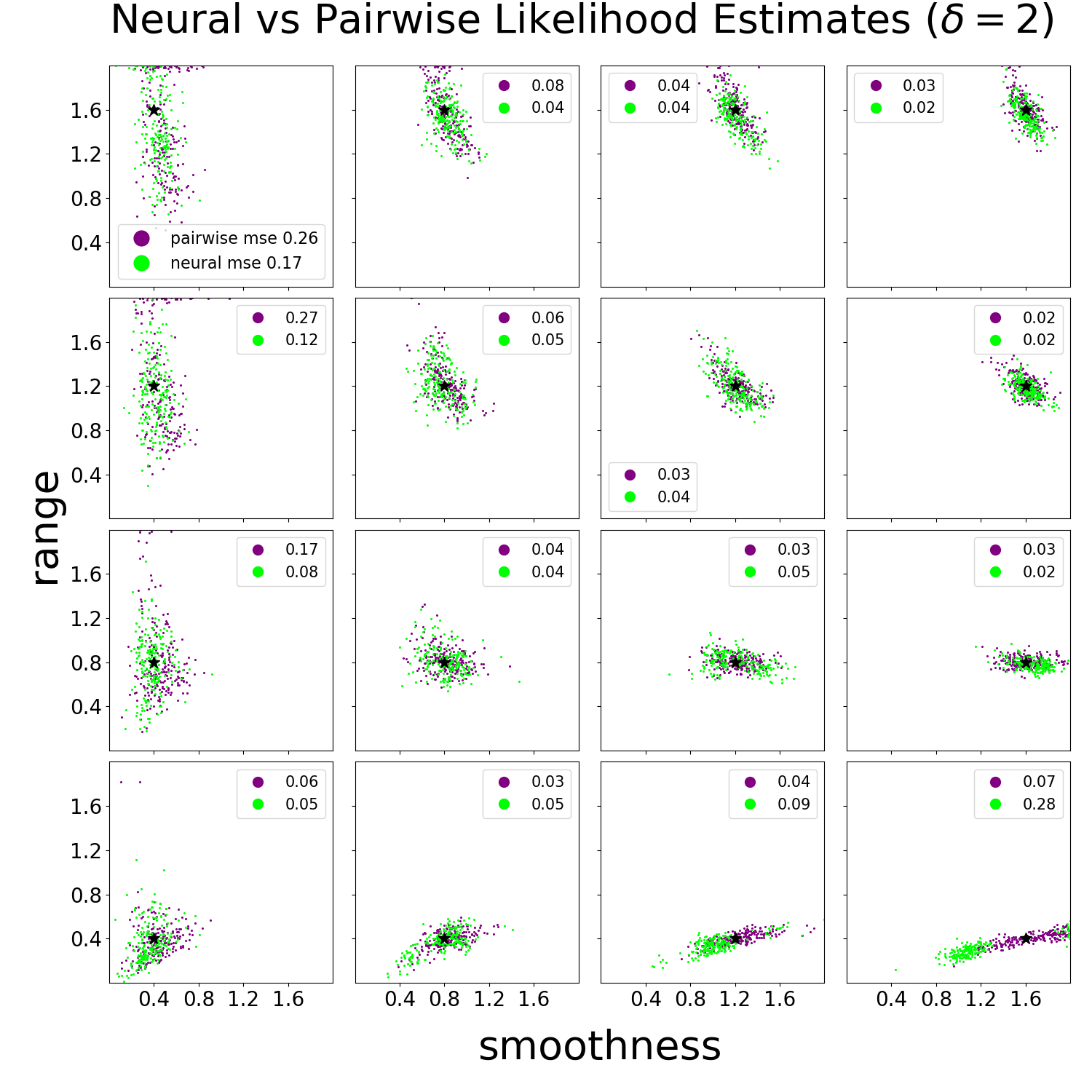}
    \caption{Parameter estimates for neural likelihood and pairwise likelihood with $\delta = 2$. Each of the $16$ plots contains the true parameter (black star) which generated the $200$ spatial field realizations and the corresponding parameter estimates for pairwise likelihood (purple) and neural likelihood (green) with mean squared error (MSE) in the legend. The true parameter increases in range from bottom to top and in smoothness from left to right.}
    \label{fig:brparameters}
\end{figure}

\paragraph{\textit{Evaluation: Parameter Estimation}}
In Figure~\ref{fig:brparameters}, we compare parameter estimates for neural and pairwise likelihood for cut-off $\delta=2$, the optimal cut-off in terms of pairwise parameter estimation. To determine the optimal cut-off, we varied $\delta$ and observed that the accuracy of pairwise estimates increases as $\delta$ increases from $0$ to approximately $2$ and then decreases as $\delta$ increases beyond $2$ as shown in Table~\ref{Tab:parampwl}.  See Section \ref{supplement:additionalresults} of the Supplementary Material for pairwise estimates with different cut-offs $\delta$. From Figure~\ref{fig:brparameters}, we observe that the neural and pairwise estimates are similar in terms of accuracy and behavior except in certain areas of the evaluation parameter space $\Theta$. For large smoothness and small range, the pairwise estimates are significantly more accurate than the neural estimates because the pairwise estimates are less biased and have less variance. For small smoothness and large range, the neural estimates are more accurate than the pairwise estimates.

\begin{table}[!t]
\caption{Root mean squared error (rmse), mean absolute error (mae), and median of the median absolute error (mmae) for all parameter estimates (200 estimates per parameter on a $9\times 9$ grid over $\Theta$) for a Brown--Resnick process for pairwise likelihood with different distance cut-offs $\delta$ and neural likelihood. The best results for each metric are in~bold.}
\small
\begin{center}
\begin{tabular}{ |p{5cm} p{1cm} p{1cm}  p{1cm}|}
 \hline
 
 Type of surface & rmse & mae & mmae\\
 \hline
 pairwise likelihood ($\delta = 1$) & 0.51 & 0.74 & 0.65 \\
 pairwise likelihood ($\delta = 2$) & 0.25 & $\bm{0.30}$ & $\bm{0.20}$\\
 pairwise likelihood ($\delta = 5$) & 0.28 & 0.37 & 0.25\\
 neural likelihood & $\bm{0.24}$ & $\bm{0.30}$ & $\bm{0.20}$\\
 \hline
\end{tabular}
\end{center}
\label{Tab:parampwl}
\end{table}

\paragraph{\textit{Evaluation: Approximate Confidence Regions}}
As shown in Figure~\ref{fig:brsurfaces}, the approximate confidence regions for pairwise likelihood increase in size after adjustment. The neural and pairwise confidence regions have remarkably different shapes and sizes no matter whether the pairwise likelihood is adjusted or not. As in the Gaussian process case, calibration increases the size while preserving the shape and location of the neural likelihood confidence region.

\paragraph{\textit{Evaluation: Empirical Coverage and Confidence Region Area}}

\begin{figure}[!t]
    \centering
    \includegraphics[scale = .27]{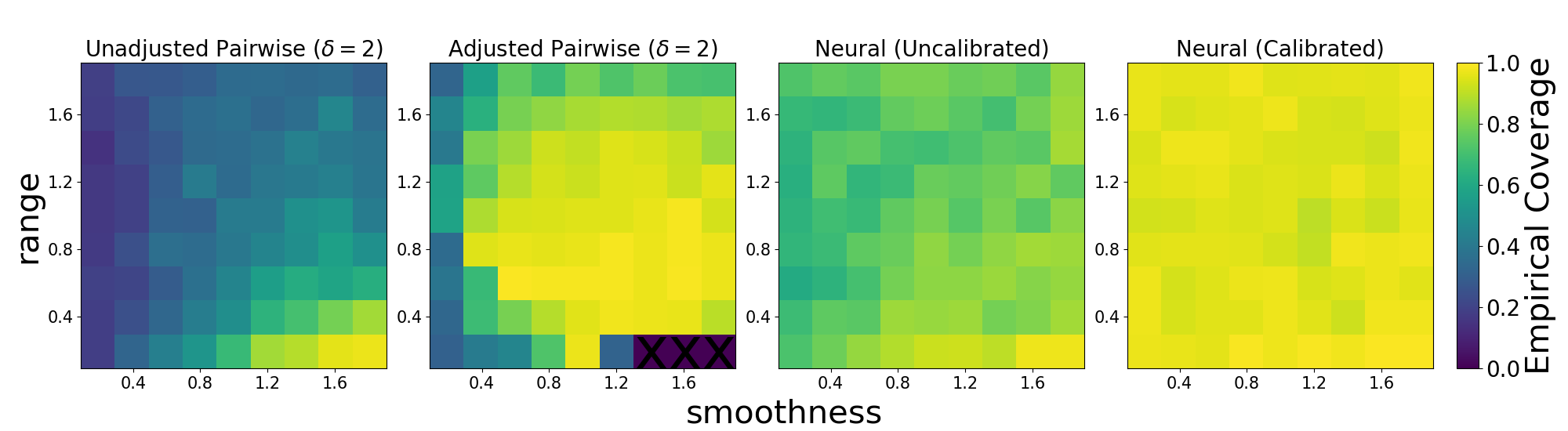}
    \caption{Empirical coverage for $95\%$ approximate confidence regions for unadjusted pairwise likelihood with distance cut-off $\delta = 2$ (far left) and adjusted pairwise likelihood with $\delta =2$ (center left) and neural likelihood before calibration (center right) and after calibration (far right).}
    \label{fig:brempiricalcoverage}
\end{figure}

\begin{figure}[!t]
    \centering
    \includegraphics[scale = .27]{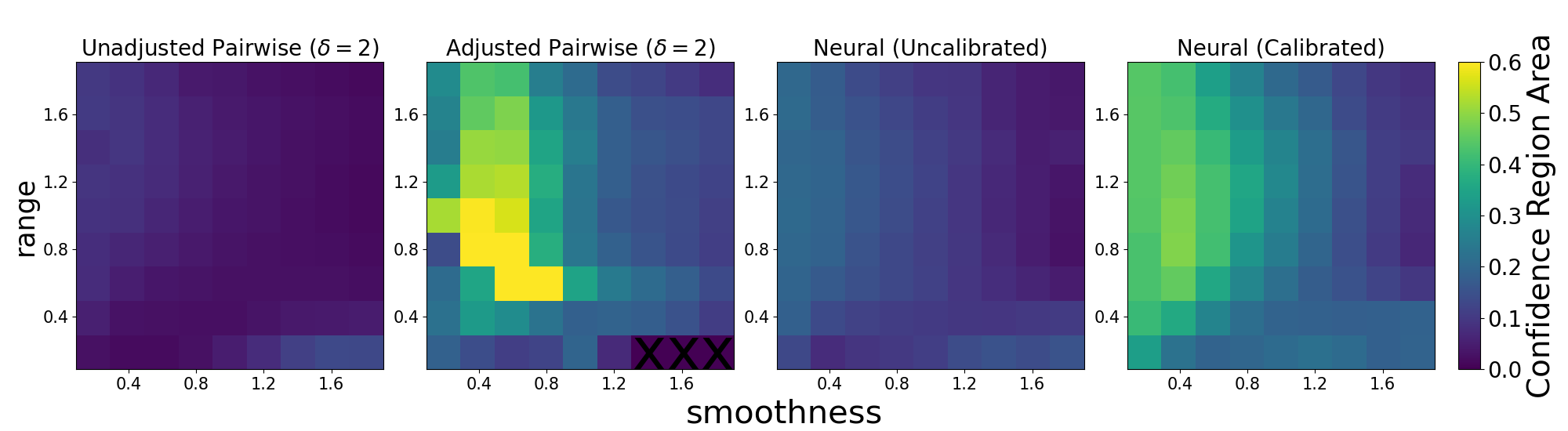}
    \caption{ $95\%$ approximate confidence region area for unadjusted pairwise likelihood with a distance cut-off of $\delta = 2$ (far left) and adjusted pairwise likelihood with $\delta = 2$ (center left) and neural likelihood before calibration (center right) and after calibration (far right).}
    \label{fig:brconfidenceregionarea}
\end{figure}

After adjustment, confidence region area and thus empirical coverage increase for the $95\%$ approximate confidence regions constructed from pairwise likelihood surfaces as shown in Figures~\ref{fig:brconfidenceregionarea} and \ref{fig:brempiricalcoverage} respectively. Before adjustment, the confidence regions are unrealistically small, and thus, empirical coverage is low. After adjustment, the confidence regions are reasonably sized and achieve decent empirical coverage as a result. Yet, empirical coverage  suffers near the boundaries of the parameter space $\Theta$ for adjusted pairwise likelihood and even fails for a few parameters near the boundary (indicated by crosses in Figures~\ref{fig:brempiricalcoverage} and \ref{fig:brconfidenceregionarea}). See Section \ref{supplement:adjustment} of the Supplementary Material for details on this failure at the boundary.

As in the Gaussian process case, calibration increases confidence region area for the neural likelihood and thus increases empirical coverage close to the intended $95\%$ coverage. Empirical coverage for neural likelihood is uniform over the evaluation parameter space $\Theta$ in contrast to both variants of pairwise likelihood in which coverage varies. 
Where adjusted pairwise and calibrated neural likelihood have comparable coverage, the former tends to have larger confidence region area. To conclude, neural likelihood provides better empirical coverage than both unadjusted and adjusted pairwise likelihood for distance cut-off $\delta=\nobreak2$ while keeping confidence region area relatively small and coverage across $\Theta$ relatively uniform.

\paragraph{\textit{Evaluation: Timing Study}}
As mentioned earlier in this section, we used the function \texttt{nllh} from the \texttt{fitmaxstab} function to evaluate the pairwise likelihood surfaces. Since the \texttt{fitmaxstab} function involves finding the parameters that best fit the spatial field via L-BFGS, we limited the number of optimization iterations to zero to only count the time necessary to construct the pairwise likelihood surface. We used parallel processing with the same processor in the Gaussian process case study to evaluate the function \texttt{nllh} at each parameter of the $40\times 40$ grid over the evaluation parameter space $\Theta$. Importantly, our timing study only involves unadjusted pairwise likelihood surfaces. The time to produce adjusted pairwise likelihood surfaces is similar if the computation of the linear transformation for the adjustment is not considered. Additionally, as in the Gaussian process case, the time to train the neural network is not included in the recorded times to evaluate the neural likelihood surfaces. The time to train the neural network in the Brown--Resnick case is similar to the Gaussian process case.

As the distance cut-off $\delta$ increases, the time to construct the corresponding pairwise likelihood surface should increase and indeed does increase as shown in Table~\ref{Tab:brtime} because the number of spatial location pairs for which the bivariate likelihood is computed increases. Yet, the computation time only increases slightly as $\delta$ increases. Depending on $\delta$, the vectorized neural likelihood method is approximately two to three times faster than pairwise likelihood. However, the unvectorized method is at least twice as slow as computing the pairwise likelihood surface. Thus, the efficiency of neural likelihood surfaces is due to the ability to evaluate the CNN at multiple inputs simultaneously.

\begin{table}[!t]
\caption{Time to produce neural and unadjusted pairwise likelihood surfaces on a $40\times 40$ grid over $\Theta$ for $50$ realizations of a Brown--Resnick process on a $25\times 25$ grid on spatial domain $[-10,10]\times [-10,10]$.}
\small
\begin{center}
\begin{tabular}{ |p{6cm} p{2.5cm} p{4.5cm}| }
 \hline
 Type of surface and method & average (sec) & standard deviation (sec)\\
 \hline
 pairwise likelihood ($\delta=1$) & 5.05 & 0.34\\
 pairwise likelihood ($\delta=2$) & 5.38 & 0.27\\
 pairwise likelihood ($\delta=5$) & 5.86 & 0.28\\
 pairwise likelihood ($\delta=10$) & 7.33 & 0.16\\
 vectorized neural likelihood & 2.24 & 0.13\\
 unvectorized neural likelihood & 14.39 & 0.03\\
 \hline
\end{tabular}
\end{center}
\label{Tab:brtime}
\end{table}

\paragraph{\textit{Evaluation: Multiple Realizations}}

\begin{figure}[!t]
    \centering
    \includegraphics[scale = .32]{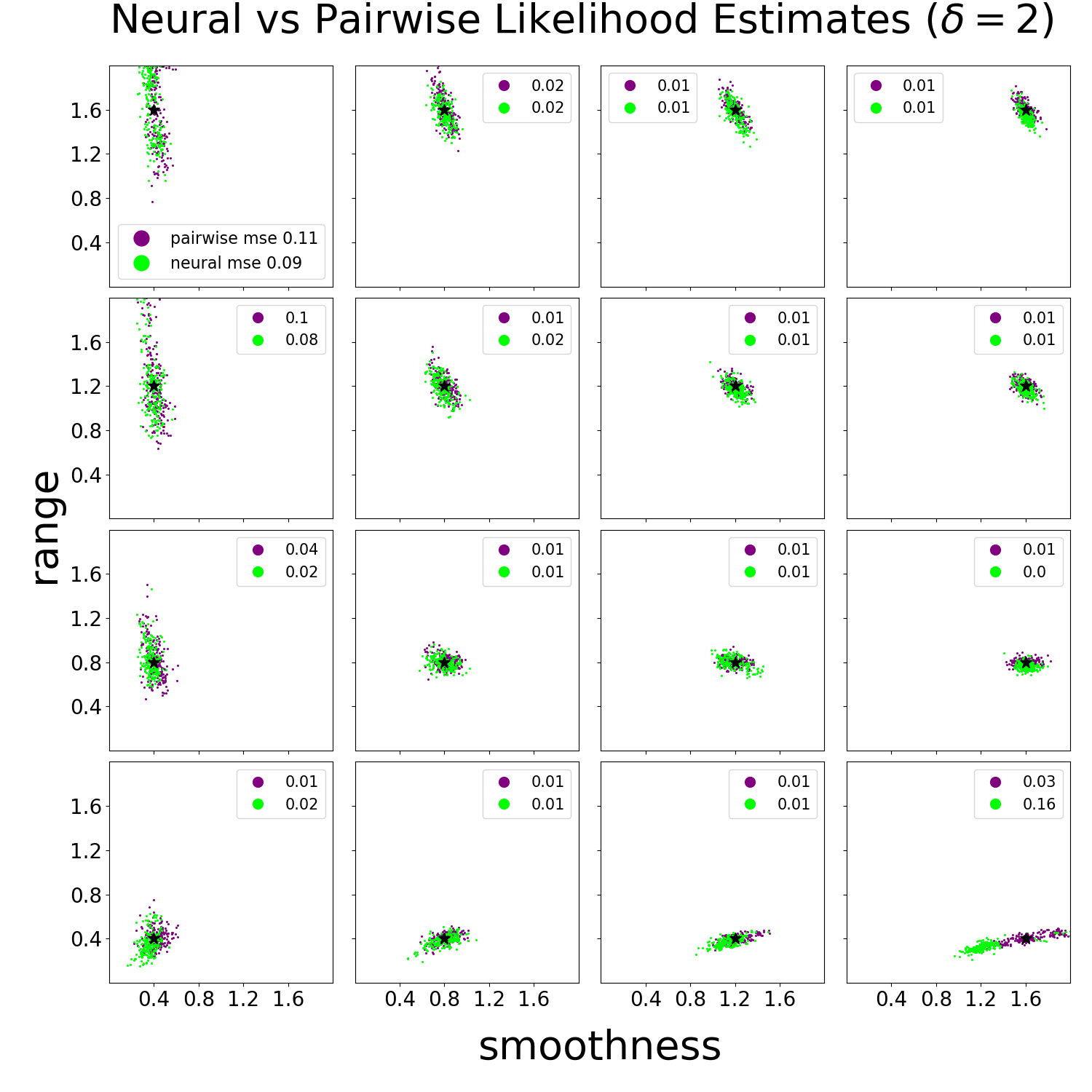}
    \caption{Parameter estimates for neural likelihood and pairwise likelihood with $\delta = 2$ in the case of 5 i.i.d.\ spatial field realizations for a Brown--Resnick process. Each of the $16$ plots contains the true parameter (black star) which generated the realizations and the corresponding parameter estimates for pairwise likelihood (purple) and neural likelihood (green) with mean squared error (MSE) in the legend. The true parameter increases in range from bottom to top and in smoothness from left to right.}
    \label{fig:brfiveparams}
\end{figure}

The parameter estimators for pairwise and neural likelihood for the five i.i.d.\ realizations case in Figure~\ref{fig:brfiveparams} are more accurate and have less variance than the equivalent for the single realization case in Figure~\ref{fig:brparameters}, as expected. As in the single realization case, the neural and pairwise estimates are similar in terms of accuracy and behavior except in certain areas of the evaluation parameter space $\Theta$. For large smoothness and small range, the pairwise estimates are significantly more accurate than the neural estimates. Yet, for small smoothness and large range, the neural estimates are slightly more accurate than the pairwise estimates.

\paragraph{\textit{Summary of Results}}

In this case study, we demonstrated that in terms of parameter estimation and uncertainty quantification, our method of learning the likelihood has significant advantages over pairwise likelihood, a well-established approximation for the intractable exact likelihood. The neural likelihood parameter estimates are generally comparable or significantly better than the pairwise likelihood parameter estimates depending on the distance cut-off $\delta$. Yet, there are some cases for which the pairwise estimates are more accurate depending on the location in the parameter space $\Theta$. This discrepancy may be due to either the CNN not sufficiently learning the likelihood in this particular region or it could be that pairwise likelihood is truly better than exact likelihood in terms of parameter estimation in this region. The approximate confidence regions for neural likelihood provide better uncertainty quantification than pairwise likelihood whether adjusted or not: the neural confidence regions achieve empirical coverage comparable to the intended coverage with sufficiently small confidence region area. From the evaluation of neural likelihood surfaces, confidence regions, and empirical coverage before and after calibration, it is clear that calibration is essential to our method in order to achieve the intended coverage level of the approximate confidence regions. Finally, the neural likelihood surfaces are significantly faster to evaluate than pairwise likelihood surfaces because neural networks are fast to evaluate and our CNN, once trained, is amortized.

\section{Discussion and Conclusions}
\label{sec:discuss}
In this paper, we have proposed a new method to learn the likelihood function of spatial processes using a specifically designed classification task. This classification task involves generating simulated data from the spatial process to construct two classes which consist of pairs of dependent and independent spatial fields $\bm{y}$ and parameters $\bm{\theta}$. Due to the construction of the classes, the resulting classifier is equivalent to the likelihood up to a multiplicative constant and a known transformation. Once we calibrate the classifier, we can transform the classifier outputs to produce neural likelihood surfaces, approximate confidence regions, and parameter estimates.  

We demonstrated that the neural likelihood produces comparable results to the exact likelihood in terms of parameter estimation and uncertainty quantification for a Gaussian process, a spatial process with a computationally intensive yet tractable likelihood. However, evaluating the neural likelihood surface is faster than evaluating the exact likelihood surface by potentially orders of magnitude depending on the number of observed spatial locations. Additionally, the approach is well-suited to determining parameter estimates using a grid-based approach which ensures that the resulting estimator is close to the global maximizer of the neural likelihood. Altogether, the neural likelihood provides comparable parameter estimation and uncertainty quantification to the exact likelihood, when one is available, along with much improved computational efficiency and guarantees for finding parameter estimators close to the global maximizer.

We have provided compelling evidence that our method produces a neural likelihood which approximates well the exact likelihood of the Brown--Resnick process, a spatial process with an intractable likelihood. We compared neural likelihood to pairwise likelihood, a common approximation for the exact likelihood of this process. Pairwise likelihood has a tuning parameter, the distance cut-off $\delta$, which is hard to choose optimally, and an adjustment, which can ensure asymptotic guarantees of the maximum pairwise likelihood estimator yet is difficult to apply in practice. Due to this tuning parameter, there is often a trade-off in terms of obtaining reasonable surfaces, parameter estimates, and approximate confidence regions. Neural likelihood, on the other hand, does not have such a tuning parameter and does not suffer from this trade-off between good parameter estimation and reasonable approximate confidence regions. Adjusting pairwise likelihood can improve the quality of the surfaces and increase empirical coverage, yet it is tricky and computationally intensive to implement especially for different tuning parameters~$\delta$. In contrast, neural likelihood has no such adjustment, unless one considers calibration an adjustment, and performs better than adjusted pairwise likelihood in terms of surfaces, approximate confidence regions, empirical coverage, and confidence region area across most of the parameter space, although in certain parts of the parameter space, parameter estimation via pairwise likelihood was more accurate than neural likelihood. Additionally, neural likelihood surfaces are much quicker to evaluate than pairwise likelihood. Thus, we conclude that for this particular spatial process with an intractable likelihood, neural likelihood generally performs much better than pairwise likelihood.

We conclude with a discussion of the limitations and potential extensions of our method of learning the likelihood via classification. Currently, our method learns the likelihood over a predetermined bounded parameter space which is a common assumption in contemporary simulation-based inference \citep[e.g.,][]{Dalmasso} since simulating training parameters from an unbounded space is not feasible in practice. An important question for future work is to address the misspecification situation where the neural likelihood is evaluated for a realization $\bm{y}$ of the spatial process generated by a parameter $\bm{\theta}$ outside the parameter space used to train the classifier. With active learning \citep[646--647]{Murphy}, we may be able to adaptively extend the parameter space during training in such a way that even in the misspecified case we eventually include with high probability the region that contains the true parameter $\bm{\theta}$.

Another limitation is the assumption that the spatial observations are on a fully observed regular grid of fixed size. This motivates two extensions that we plan to address in future work: First, it would be important to extend the method to partially observed grids which can probably be done by considering the unobserved locations on the grid as latent variables. Second, there are many significant examples of irregular spatial data which our method cannot currently handle. One such example is Argo float data \citep{Wong,Kuusela2018} consisting of irregularly sampled ocean temperature and salinity profiles throughout the global ocean. In the future, we would like to extend neural likelihood to such irregular spatial data, which is a nontrivial extension since the neural network and its training would need to be adapted to handle irregular inputs. One possible method of extension is graph convolutional neural networks which \citet{SainsburyDale_2023b} utilized to extend neural estimation to irregular spatial data.

Another limitation of this work is our focus on low-dimensional parameter spaces. Likelihood-free inference methods similar to the one considered here have been shown to work with at least up to ten parameters \citep[pp. 43--48]{Dalmasso2021}. However, as the parameter space grows in dimension, sampling from this space to generate a sufficient amount of training data to adequately learn the likelihood becomes exponentially more computationally demanding. This training issue may potentially be addressed by finding ways to adaptively sample the parameter space so that the training sample is concentrated in areas of high likelihood. Additionally, once the neural likelihood is obtained over a high-dimensional parameter space, the grid-based approach to parameter estimation and confidence region construction will become computationally expensive. A possible solution is a gradient-based approach in which the neural likelihood gradient and Hessian are computed using automatic differentiation.

Another important topic for future work will be to investigate the robustness  of our method to input data whose distribution deviates slightly from the stochastic model used to train the network \citep{Drenkow}. Real-data applications may require developing techniques for addressing this distribution shift if it turns out to have a substantial impact on the performance of the network.

Finally, in this paper, we have only presented our method in terms of learning the likelihood for spatial processes. Yet, this method of learning the likelihood can extend with simple modifications beyond spatial processes to other complex statistical models. The only true requirement for the statistical model of interest is the ability to easily simulate observations from it. Altogether, this method may eventually enable efficient and accurate likelihood-based parameter estimation and uncertainty quantification for a broad range of statistical models where only approximate or computationally inefficient inference has previously been possible.

\section*{Competing interests}
No competing interests are declared.



\section*{Data availability statement}

The code used to produce these results is publicly available at \url{https://github.com/jmwalchessen/neural_likelihood}. The simulation experiments in this paper can be reproduced with the given code.

\section*{Acknowledgments}
We are grateful to the members of the CMU Statistical Methods for the Physical Sciences (STAMPS) Research Group for insightful discussions and feedback throughout this work. We would also like to thank Raphael Huser and L\'{e}o Belzile for helpful discussions about spatial extremes. We would like to acknowledge Microsoft for providing Azure computing resources for this work. J.W.\ and M.K.\ were supported in part by NSF grants DMS-2053804 and PHY-2020295, NOAA grant NA21OAR4310258 and a grant from C3.ai Digital Transformation Institute.


\bibliographystyle{Chicago} 
\bibliography{cas-refs}

\clearpage

\setcounter{page}{1}
\beginsupplement
\section{Supplementary Material}
\label{supplement}

\subsection{Background Information on Convolutional Neural Networks}
\label{supplement:cnn}

The following offers a brief overview of CNNs as used in our methodology. For those interested in a more detailed description, we refer the reader to \citet{Lecun}, the original paper introducing CNNs, and Chapter~9 of \citet{Goodfellow}. A CNN is a type of neural network which takes as input a matrix of fixed size and uses convolutions to extract spatial patterns between matrix entries pertinent to the task in question \citep[pp.~329--341]{Goodfellow}. Specifically, filters in each convolutional layer of the CNN extract different spatial patterns that may be of use in the learning task. Due to their ability to extract spatial patterns, CNNs are often used to process visual data such as photos in which pixels can be depicted as constituting a fixed-size matrix. As such, CNNs are well suited to dealing with the spatial field input which is effectively an image. Outputs of a CNN can vary from a scalar in the case of binary classification to a multidimensional vector in the case of multi-class classification or even another matrix in the case of image-on-image regression \citep[pp.~352--353]{Goodfellow}.

In a CNN, convolutional layers are interspersed with pooling layers in order to distill the spatial information contained in an image or other matrix-type object to the most important spatial patterns. Pooling layers reduce the output size of the previous layer by combining multiple neurons into a single neuron in the next layer \citep[pp.~335--339]{Goodfellow}. 
The output of the convolutional and pooling layers is a reduced-size matrix containing spatial patterns of interest to the learning task.

In the case of classification, this reduced matrix can be flattened to a manageable-size vector which is then processed in the second part of the network---a collection of fully connected layers with multiple hidden layers ending with an output layer. For binary classification, the output layer is either a binary classification output or a class probability, a real value between zero and one, depending on the activation function as well as the loss function. 

\subsection{Training Details}
\label{supplement:train}

\subsubsection{Architecture}
Table~\ref{Tab:architecture} shows the specific CNN architecture used in both case studies in Section~\ref{sec:casestudies}. 
\begin{table}[H]
\caption{CNN architecture}
\begin{center}
\begin{tabular}[h!]{|c c c c c c|} 
\hline 
 layer type & output shape & filters & kernel size & activation & weights
\\ [0.1ex]
\hline 
2D convolution & [ -, 23, 23, 128] & 128 & $3\times 3$ & ReLU & 1280\\
2D convolution & [ -, 10, 10, 128] & 128 & $3\times 3$ & ReLU & 147584\\
2D convolution & [ -, 3, 3, 16] & 16 & $3\times 3$ & ReLU & 18448\\
flatten and concatenate & [-, 66] &  & & & 0\\
Dense & [-,64] & & & ReLU & 4288\\
Dense & [-,16] & & & ReLU & 1040\\
Dense & [-,8] & & & ReLU & 136\\
Dense & [-,2] &  & & ReLU & 18\\
[.1ex] 
\hline
\end{tabular}
\end{center}
\label{Tab:architecture}
\end{table}
\subsubsection{Gaussian Process}
\label{supplement:traingp}

When experimenting with different batch sizes and learning rate schedules, we observed that as batch size decreases, overfitting tends to increase for a fixed learning rate schedule. Decreasing initial learning rate reduces but does not eliminate overfitting. We did not observe a similar tendency for overfitting during training for the Brown--Resnick process and speculate that the CNN architecture might be too complex for learning the relatively simple likelihood of a Gaussian process unless care is taken when selecting batch size and learning rate schedule.

\subsubsection{Brown--Resnick Process}
\label{supplement:trainbr}

We had some difficulty training the CNN for various combinations of learning rate schedule and batch size. The training and validation loss would generally plateau after a few epochs. Simply training the CNN with the same batch size and learning rate schedule for various initializations of the weights eventually produced a model in which the training and validation loss did not plateau early in training. The selected model for which plateauing did not occur was trained using a batch size of $50$ and a learning rate schedule in which the learning rate started at $0.002$ for the first five epochs and decreased by a multiplicative factor of $e^{-0.1}$ for each of the fifteen epochs after the first five.

As in the Gaussian process case, the evaluation parameter space is $\Theta = (0,2)\times (0,2)$. We experimented with a training parameter space $\tilde{\Theta}=(0,2)\times (0,2.5)$ in which the domain of the smoothness parameter $\nu$ is not extended because this parameter is bounded from zero to two as described in Section \ref{sec:brdescription}. Yet, extending $\tilde{\Theta}$ beyond the evaluation parameter space $\Theta$ caused problems during training. The validation loss fluctuated while the training loss smoothly decreased as the number of epochs increased. In contrast, the validation loss smoothly decreased with the training loss for the CNN trained on data from $\Theta=(0,2)\times (0,2)$. Thus, the model we selected was trained on data from the training parameter space $\tilde{\Theta}=(0,2)\times (0,2)=\Theta$ with the batch size and learning rate schedule described earlier.

\subsection{More Description of Brown--Resnick Process and its Approximate Likelihood}
\label{supplement:brdescription}

In general, we can construct a max-stable process from other stochastic processes in the following way. First, consider the points $\eta_{i}$ of a positive Poisson point process with intensity function $\d\Lambda(\eta) = \eta^{-2} \d \eta$ and i.i.d.\ realizations $W_{i}(\bm{s})$ of a non-negative stochastic process with a mean of one. For independent realizations of the Poisson process and the non-negative process, a max-stable process has the following form:
\begin{equation}
    Z(\bm{s}) = \max_{i > 0} \eta_{i} W_{i}(\bm{s}) \textrm{ for a location } \bm{s} \textrm{ on a domain }\mathcal{D}\subset \mathbb{R}^{d}
\label{eqn:maxstable}
\end{equation}
with marginal $P(Z(\bm{s})\leq z) = \exp(-\frac{1}{z})$, the unit Fr\'echet distribution \citep{Kabluchko}. 

Depending on what non-negative stochastic process we select for $W_{i}(\bm{s})$, we obtain different max-stable processes \citep{Kabluchko}. We can construct the Brown--Resnick process using \eqref{eqn:maxstable} with $W_i(\bm{s})=\exp(\epsilon_i(\bm{s}) - \gamma(\bm{s}))$, where $\epsilon_i(\bm{s})$ are realizations of an intrinsically stationary Gaussian process such that $\epsilon_{i}(\bm{s})=0$ almost surely with semivariogram $\gamma(\bm{h})=(\frac{\norm{\bm{h}}}{\lambda})^{\nu}$ where $\bm{h}$ is the spatial separation between two locations, $\lambda \in \mathbb{R}^{+}$ is a range parameter, and $\nu\in (0,2]$ is a smoothness parameter \citep{Castruccio}. As such, the parameter space $\Theta$ for which we are interested in learning the likelihood is a bounded subset of $\mathbb{R}^{+}\times (0,2]$.

The density is
\begin{equation}
f(z_{1},\dots, z_{n}) = \exp\big(-V(z_{1},\dots,z_{n})\big) \sum_{\mathcal{P} \in \mathcal{P}_{z}} \prod_{S\in \mathcal{P}} -V_{S}(z_{1},\dots,z_{n})
\label{eqn:jointpdf}
\end{equation}
where $\mathcal{P}_{z}$ is the set of all partitions $\mathcal{P}
$ of the values $z_{1},\dots,z_{n}$ and $V_{S}$ is the partial derivative of 
\begin{equation}
    V(z_{1},\dots,z_{n})=E(\max \{\frac{W(\bm{s}_{1})}{z_{1}},\dots, \frac{W(\bm{s}_{n})}{z_{n}}\})
\end{equation}
with respect to the values $z_{1},\dots,z_{n}$ indexed by $S\in \mathcal{P}$ \citep{Castruccio, Huser}. The number of terms in the summation in \eqref{eqn:jointpdf} is $\abs{P_{z}}=B_{n}$, the Bell number for $n$, which grows more than exponentially \citep{Castruccio}.

Since the number of spatial locations is generally large in practice, the computation of the likelihood for a Brown--Resnick process is intractable in many practical cases. Yet, in the bivariate case of only two spatial locations, the likelihood has a simple closed form. From \eqref{eqn:jointpdf}, the bivariate log likelihood is 
\begin{equation}
\ell(\lambda, \nu) = \log\Big(V_{1}(z_{1},z_{2})V_{2}(z_{1},z_{2}) - V_{12}(z_{1},z_{2})\Big) - V(z_{1},z_{2})
\label{eqref:bivariatefulllikelihood}
\end{equation}
in which V and its partial derivatives $V_{1},V_{2}$ and $V_{12}$ are tractable \citep{Huser}. Using the bivariate case of the full likelihood, we can provide an approximation for the full likelihood of a Brown--Resnick process called pairwise likelihood, a form of composite likelihood. \citep{padoan2010likelihood}. 

In pairwise log likelihood, the summands are the bivariate likelihoods between two spatial locations $\bm{s}_{j_{1}}$ and $\bm{s}_{j_{2}}$ for $j_{1},j_{2}\in [n]$, and involve only select pairs of spatial locations $(\bm{s}_{j_{1}},  \bm{s}_{j_{2}})$ because even a summation over all $\frac{n(n-1)}{2}$ pairs of spatial locations is computationally intensive for a large number of spatial locations. The pairwise log likelihood has the following form \citep{Huser, Castruccio}:

\begin{equation}
\begin{aligned}
\log\Big(\mathcal{L}_{\textrm{approx}}\big((\lambda, \nu) \mid \bm{z}\big)\Big)& = \sum_{j_{2}>j_{1}} \sum_{j_{1}=1}^{n} w_{j_{1},j_{2}}\bigg[\log\Big(V_{1}(z_{j_{1}}, z_{j_{2}})V_{2}(z_{j_{1}},z_{j_{2}}) \\
&- V_{12}(z_{j_{1}},z_{j_{2}})\Big) - V(z_{j_{1}},z_{j_{2}})\bigg]
\end{aligned}
\end{equation} where the weights $w_{j_{1},j_{2}}$ are between zero and one depending on whether the bivariate likelihood for the pair of spatial locations $(\bm{s}_{j_{1}},\bm{s}_{j_{2}})$ is included in the summation. Inclusion is based on a criterion such as the following:
\begin{equation}
    w_{j_{1},j_{2}} = 
    \begin{cases}
    1, & \textrm{ if } \norm{\bm{s}_{j_{1}} - \bm{s}_{j_{2}}} \leq \delta,\\
    0, & \textrm{ otherwise}.
    \end{cases}
\label{eqn:weightpair}
\end{equation} Since nearby locations contain the most information about a given location, the selection criterion generally is determined by the distance between spatial locations. In \eqref{eqn:weightpair}, all pairs of observations whose locations are within a certain cut-off distance $\delta$ of each other are included.

When computing the pairwise likelihood, the cut-off distance $\delta$ is a tuning parameter that must be appropriately selected in order to obtain reasonable results. In practice, if the cut-off distance is too small or too large, the maximum pairwise likelihood estimates can be highly inaccurate, and the pairwise likelihood surfaces tend to be uninformative as shown in Section \ref{supplement:additionalresults}. In contrast, neural likelihood does not involve such tuning parameters.

Additionally, the asymptotic behavior of the maximum likelihood estimator and, subsequently, the likelihood ratio statistic is different for pairwise likelihood. Under certain conditions, the maximum likelihood estimator for pairwise likelihood has the following asymptotic distribution \citep{padoan2010likelihood}:
\begin{equation}
\hat{\bm{\theta}}_{\textrm{pwl}} \sim \mathcal{N}(\bm{\theta}^{*}, H(\bm{\theta}^{*})^{-1}J(\bm{\theta}^{*})H(\bm{\theta}^{*})^{-1}),
\label{eq:pairwise_likelihood_mle}
\end{equation}
where $H(\bm{\theta}^{*})=-\mathbb{E}\big(\frac{\partial^{2} \ell_{\textrm{pwl}}(\bm{\theta})}{\partial \bm{\theta} \partial \bm{\theta}^{\intercal}}\big)\Big|_{\bm{\theta}=\bm{\theta^{*}}}$ is the negative expectation of the Hessian of the log-pairwise likelihood at $\bm{\theta}^{*}$ and $J(\bm{\theta}^{*})=\mathrm{Cov}(\nabla \ell_{\textrm{pwl}}(\bm{\theta}^{*}))$ is the covariance matrix of the gradient. For the full likelihood, the maximum likelihood estimator, on the other hand, has the following asymptotic distribution \citep{Davison_2003}:
\begin{equation}
\hat{\bm{\theta}}_{\textrm{MLE}} \sim \mathcal{N}(\bm{\theta}^{*}, I(\bm{\theta}^{*})^{-1}),
\label{eq:full_likelihood_mle}
\end{equation}
where $I(\bm{\theta}^{*})=-\mathbb{E}\Big(\frac{\partial^{2} \ell(\bm{\theta})}{\partial \bm{\theta} \partial \bm{\theta}^{\intercal}}\Big)\Big|_{\bm{\theta}=\bm{\theta^{*}}}$ is the Fisher information for the full likelihood. The difference between \eqref{eq:pairwise_likelihood_mle} and $\eqref{eq:full_likelihood_mle}$ is the covariance of the asymptotic normal distribution of the estimators. The asymptotic covariance matrix in \eqref{eq:pairwise_likelihood_mle} does not match the curvature of the pairwise likelihood surface at the true parameter~$\bm{\theta}^{*}$ which affects the asymptotic distribution of the likelihood ratio statistic. 

To enable asymptotic inference, \citet{Chandler_2007} proposed adjusting the pairwise likelihood such that the asymptotic distribution of the adjusted pairwise likelihood ratio statistic is the same as that of the full likelihood. This method involves adjusting the parameters that are evaluated by the pairwise likelihood function via an affine transformation. The resulting adjusted log-pairwise likelihood $\ell_{\textrm{apwl}}$ is given by \citep{Chandler_2007}
\begin{equation}
\ell_{\textrm{apwl}}(\bm{\theta})=\ell_{\textrm{pwl}}(\tilde{\bm{\theta}}), \quad \quad \tilde{\bm{\theta}} = \hat{\bm{\theta}}_{\textrm{pwl}}+C(\bm{\theta}^{*})(\bm{\theta}-\hat{\bm{\theta}}_{\textrm{pwl}}),
\label{eq:adjustedpairwise}
\end{equation}
where $C(\bm{\theta}^{*})$ is a square matrix that has the effect of adjusting the curvature of the pairwise likelihood surface at $\hat{\bm{\theta}}_{\textrm{pwl}}$. Specifically, $C(\bm{\theta}^{*})$ is formed from $H(\bm{\theta}^{*})$, the unadjusted curvature, and $H_{\textrm{adj}}(\bm{\theta}^{*})=H(\bm{\theta}^{*})J(\bm{\theta}^{*})^{-1}H(\bm{\theta}^{*})$, the intended curvature, via
\begin{equation}
C(\bm{\theta}^{*}) = M(\bm{\theta}^{*})^{-1} M_{\textrm{adj}}(\bm{\theta}^{*}),
\label{eq:Cformula}
\end{equation}
such that $M(\bm{\theta}^{*})$ and $M_{\textrm{adj}}(\bm{\theta}^{*})$ are matrix square roots of $H(\bm{\theta}^{*})$ and $H_{\textrm{adj}}(\bm{\theta}^{*})$, respectively. By construction of $C(\bm{\theta}^{*})$, the adjusted pairwise likelihood has the required curvature at its maximizer. Consequently, the adjusted pairwise likelihood ratio statistic has an asymptotic chi-squared distribution \citep{Chandler_2007}. See Section \ref{supplement:adjustment} for further technical details on adjusting the pairwise likelihood.

\subsection{Technical Details for Adjusting the Pairwise Likelihood}
\label{supplement:adjustment}

Section \ref{supplement:brdescription} provides a broad overview of how to adjust the pairwise likelihood surface using the linear transformation $C(\bm{\theta}^{*})$ of the parameters at which we evaluate the pairwise likelihood in \eqref{eq:adjustedpairwise}. In this section, we focus on the technical details of computing $C(\bm{\theta}^{*})$ in which $\bm{\theta}^{*}$ is the true parameter that generated the spatial field realizations of interest. The linear transformation $C(\bm{\theta}^{*})$ is formed from the negative expectation of the Hessian $H(\bm{\theta}^{*})$ and the covariance of the gradient $J(\bm{\theta}^{*})$. As such, we first need to compute the Hessian $D(\bm{\theta}^{*} \mid \bm{y}) = \frac{\partial^{2} \ell_{\textrm{pwl}}(\bm{\theta} \mid \bm{y})}{\partial \bm{\theta} \partial \bm{\theta}^{\intercal}}\big|_{\bm{\theta}=\bm{\theta^{*}}}$ and gradient $\nabla \ell_{\textrm{pwl}}(\bm{\theta}^{*} \mid \bm{y})$ for multiple realizations $\bm{y}\sim p(\bm{y} \mid \bm{\theta}^{*})$. In practice, we use finite differencing to obtain numerical approximations $\tilde{D}(\bm{\theta}^{*} \mid \bm{y})$ and $\tilde{\nabla} \ell_{\textrm{pwl}}(\bm{\theta}^{*} \mid \bm{y})$ of $D(\bm{\theta}^{*} \mid \bm{y})$ and $\nabla \ell_{\textrm{pwl}}(\bm{\theta}^{*} \mid \bm{y})$, respectively.

Specifically, for a given parameter $\bm{\theta}^{*}$ on the $9\times 9$ grid over the evaluation space $\Theta=(0,2)\times (0,2)$, we simulated $n=5000$ spatial field realizations $\bm{y}_{i}$. Then, we computed $\tilde{D}(\bm{\theta}^{*} \mid \bm{y}_{i})$ and $\tilde{\nabla} \ell_{\textrm{pwl}}(\bm{\theta}^{*} \mid \bm{y}_{i})$ for all $i\in [n]$ using finite differencing. For $\tilde{\nabla} \ell_{\textrm{pwl}}(\bm{\theta}^{*} \mid \bm{y}_{i})$, we used forward differencing with $h = 0.05$, the fineness of the $40\times 40$ grid over $\Theta$ on which we evaluate the neural and pairwise likelihood surfaces. Thus, the estimator of $J(\bm{\theta}^{*})$ is
\begin{equation}
\hat{J}(\bm{\theta}^{*}) = \frac{1}{n} \sum_{i=1}^{n} \tilde{\nabla} \ell_{\textrm{pwl}}(\bm{\theta}^{*} \mid \bm{y}_{i})\tilde{\nabla} \ell_{\textrm{pwl}}(\bm{\theta}^{*} \mid \bm{y}_{i})^\intercal.
\label{eqn:jestimator}
\end{equation}

Finite differencing for $\tilde{D}(\bm{\theta}^{*} \mid \bm{y}_{i})$ is more difficult because $H(\bm{\theta}^{*})$ must be positive definite for the entries of the square root matrix to be real. To ensure the estimator is positive definite, we compute four different finite-differenced approximations of the Hessian via upper right, lower left, lower right, and upper left finite differencing, in this order, with $h = 0.05$. After computing each difference, we check whether the given approximate Hessian is negative definite. If so, we used the Hessian in our estimation of $H(\bm{\theta}^{*})$ and halted computing the remaining finite differences. Otherwise, we continued computing finite-differenced Hessians in the given order. If none of the four finite-differenced approximations were negative definite, no Hessian for the given realization $\bm{y}_{i}$ was included in the estimation of $H(\bm{\theta}^{*})$. Thus, the estimator of $H(\bm{\theta}^{*})$ is
\begin{equation}
\hat{H}(\bm{\theta}^{*}) = -\frac{1}{\tilde{n}} \sum_{i=1}^{n} \mathbf{1}\big(\tilde{D}(\bm{\theta}^{*} \mid \bm{y}_{i}) \prec 0\big) \tilde{D}(\bm{\theta}^{*} \mid \bm{y}_{i}) \textrm{ where } \tilde{n}=\sum_{i=1}^{n} \mathbf{1}\big(\tilde{D}(\bm{\theta}^{*} \mid \bm{y}_{i}) \prec 0\big).
\label{eqn:hestimator}
\end{equation}
 
Since the parameter $\bm{\theta}^{*}$ is unknown in practice, a practitioner would instead have to use certain data-driven estimators of $\hat{J}(\bm{\theta}^{*})$ and $\hat{H}(\bm{\theta}^{*})$ that do not rely on the knowledge of $\bm{\theta}^{*}$ \citep{SpatialExtremesUserGuide}. For simplicity, in this paper, we focus on the oracle estimators $\hat{J}(\bm{\theta}^{*})$ and $\hat{H}(\bm{\theta}^{*})$ defined in Eqs.~\eqref{eqn:jestimator} and \eqref{eqn:hestimator}. As such, the adjusted pairwise likelihood surfaces, approximate confidence regions, empirical coverage, and confidence region area presented in Section~\ref{sec:brexperiment} are optimistic versions of what would be available in practice for pairwise likelihood.

In certain situations, obtaining a negative definite Hessian $\tilde{D}(\bm{\theta}^{*} \mid \bm{y}_{i})$ using our particular method of finite differencing turned out to be impossible for all $n=5000$ spatial field realizations. As a result, $\hat{H}(\bm{\theta}^{*})$ cannot be computed, and the adjustment cannot be performed. We attempted to adjust the pairwise likelihood for all parameters on the $9\times 9$ grid over the evaluation parameter space $\Theta$ for $\delta = 1,2$. Only for $\delta = 2$ were we able to compute $\hat{H}(\bm{\theta}^{*})$ for most parameters on the $9\times 9$ grid, except for three. In Figures \ref{fig:brempiricalcoverage} and \ref{fig:brconfidenceregionarea}, the empirical coverage and confidence region area for these three parameters are masked out. In the case of $\delta = 1$, the adjustment was unsuccessful for large portions of the parameter space. For this reason, we present adjusted pairwise likelihood results only for $\delta=2$ in Section~\ref{sec:brexperiment}.

With positive definite $\hat{H}(\bm{\theta}^{*})$ and $\hat{J}(\bm{\theta}^{*})$, the square roots of $\hat{H}(\bm{\theta}^{*})$ and $\hat{H}_{\textrm{adj}}(\bm{\theta}^{*})=\hat{H}(\bm{\theta}^{*})\hat{J}(\bm{\theta}^{*})^{-1}\hat{H}(\bm{\theta}^{*})$ can be computed to obtain the transformation $C(\bm{\theta}^{*})$. These square roots are not unique for $k>1$, where $k$ is the number of parameters. As such, \citet{Chandler_2007} compute the square roots using either Cholesky factorization or eigendecomposition (spectral method). We implemented both methods and found that the behavior of the resulting adjusted pairwise likelihoods is similar in the local area around $\hat{\bm{\theta}}$, yet can substantially differ in the rest of the parameter space $\Theta$. In Figures \ref{fig:brsurfaces}, \ref{fig:brempiricalcoverage} and $\ref{fig:brconfidenceregionarea}$, we show the results for Cholesky factorization which, in our experiments, had greater empirical coverage than eigendecomposition over the entire parameter space and similar confidence region area.

\subsection{Additional Results}
\label{supplement:additionalresults}

As mentioned in Section~\ref{sec:casestudytwo}, depending on the distance cut-off $\delta$ for pairwise likelihood, the resulting likelihood surfaces and parameter estimates can vary dramatically in accuracy and usefulness. Here, we show additional results to illustrate how $\delta$ affects the pairwise estimates and surfaces.

\begin{figure}[h]
\centering
        \includegraphics[scale = .4]{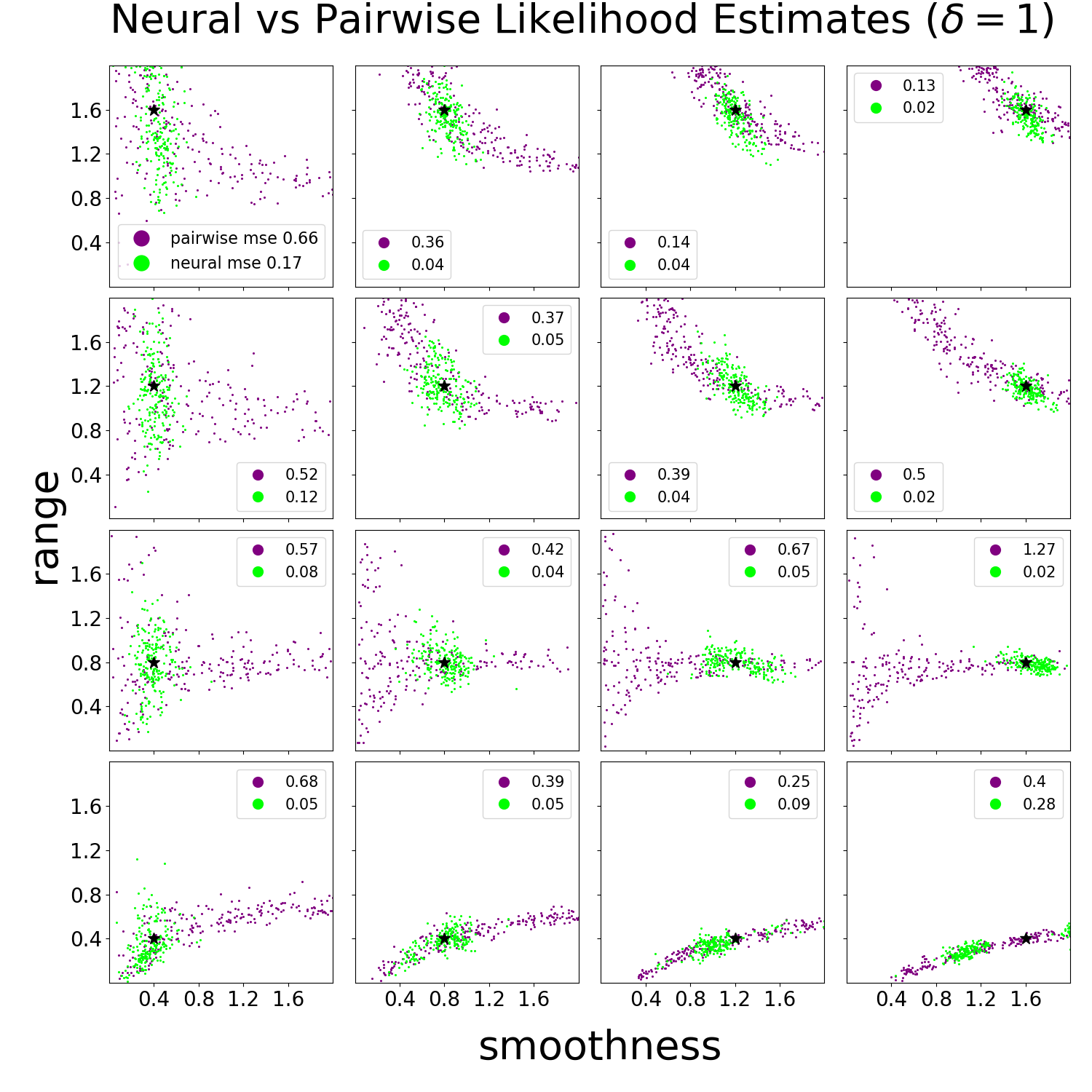}
        \caption{$4\times 4$ plot of neural and pairwise parameter estimates for $\delta = 1$. Each of the $16$ plots contains the true parameter (black star) which generated the $200$ spatial field realizations and the corresponding parameter estimates for pairwise likelihood (purple) and neural likelihood (green) with mean squared error (MSE) in the legend. The true parameter increases in range from bottom to top and in smoothness from left to right.}
        \label{fig:brparamsdelta1}
\end{figure}

\begin{figure}[h]
\centering
        \includegraphics[scale = .4]{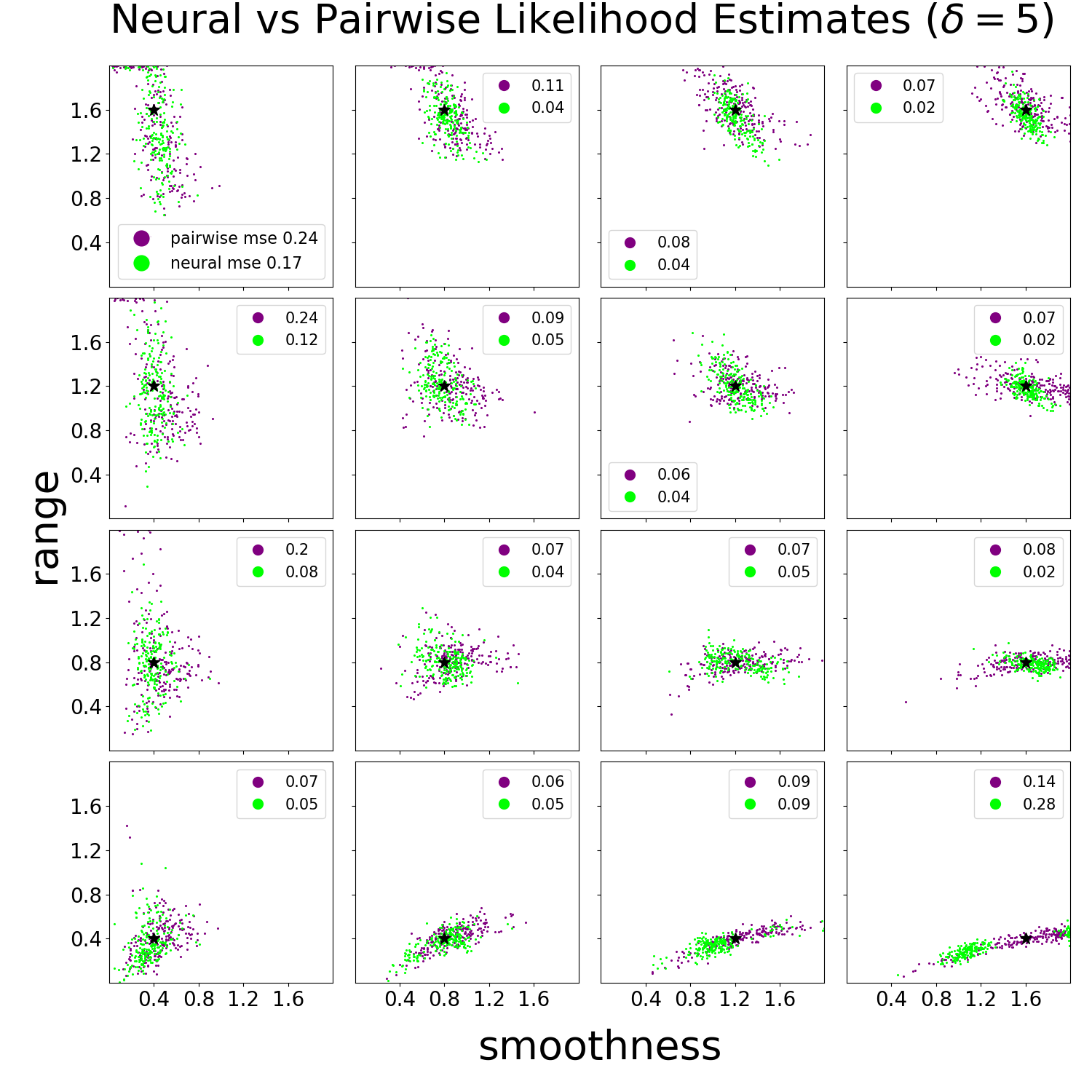}
    \caption{$4\times 4$ plot of neural and pairwise parameter estimates for $\delta = 5$. Each of the $16$ plots contains the true parameter (black star) which generated the $200$ spatial field realizations and the corresponding parameter estimates for pairwise likelihood (purple) and neural likelihood (green) with mean squared error (MSE) in the legend. The true parameter increases in range from bottom to top and in smoothness from left to right.}
    \label{fig:brparamsdelta5}
\end{figure}

\begin{figure}[h]
\centering
        \includegraphics[scale = .4]{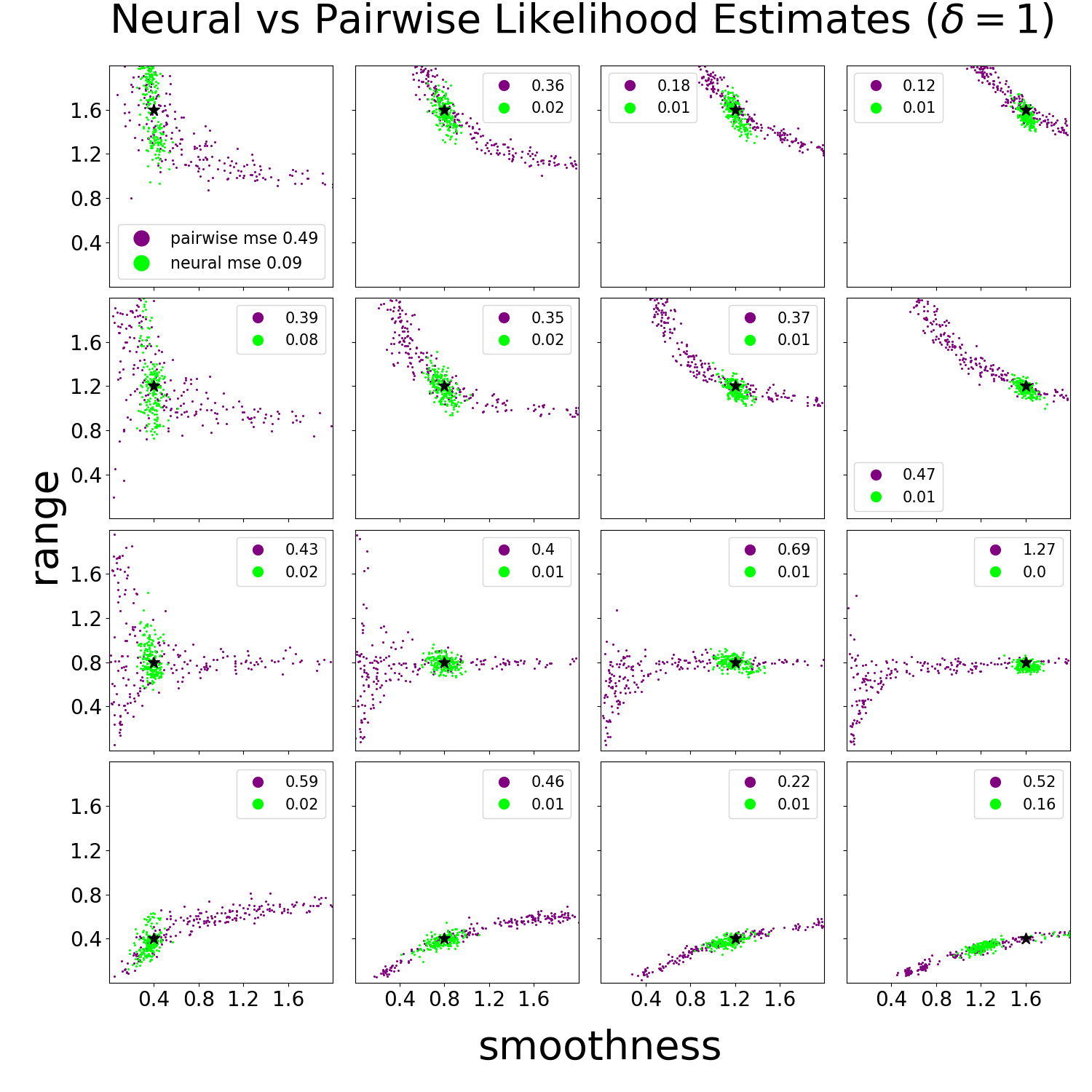}
        \caption{$4\times 4$ plot of neural and pairwise parameter estimates for $\delta = 1$ in the case of 5 i.i.d.\ spatial field realizations for a Brown--Resnick process. Each of the $16$ plots contains the true parameter (black star) which generated the realizations and the corresponding parameter estimates for pairwise likelihood (purple) and neural likelihood (green) with mean squared error (MSE) in the legend. The true parameter increases in range from bottom to top and in smoothness from left to right.}
    \label{fig:brfiveparamsdelta1}
\end{figure}

\begin{figure}[h]
\centering
        \includegraphics[scale = .4]{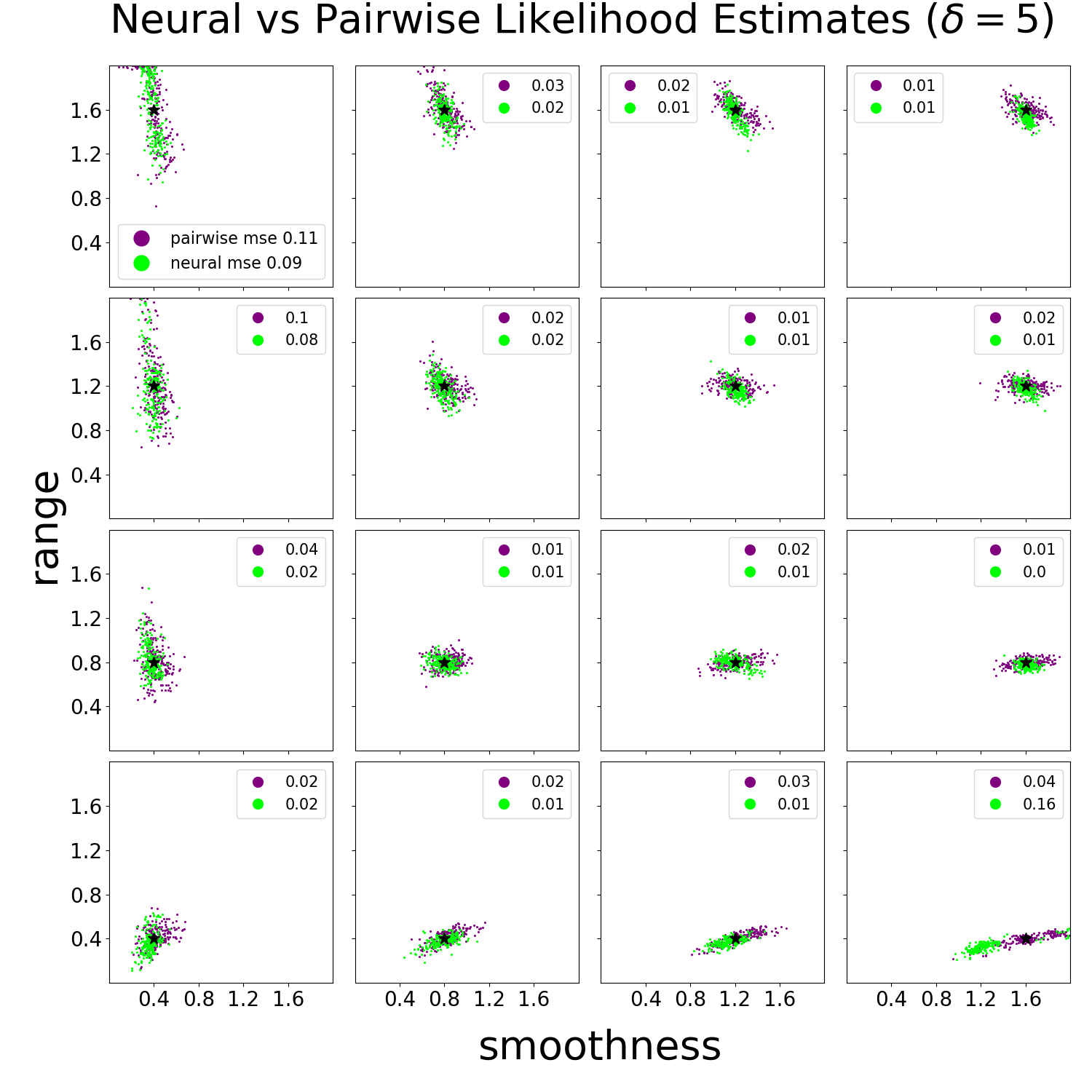}
    \caption{$4\times 4$ plot of neural and pairwise parameter estimates for $\delta = 5$ in the case of 5 i.i.d.\ spatial field realizations for a Brown--Resnick process. Each of the $16$ plots contains the true parameter (black star) which generated the realizations and the corresponding parameter estimates for pairwise likelihood (purple) and neural likelihood (green) with mean squared error (MSE) in the legend. The true parameter increases in range from bottom to top and in smoothness from left to right.}
    \label{fig:brfiveparamsdelta5}
\end{figure}

Figures~\ref{fig:brparamsdelta1} and \ref{fig:brparamsdelta5} show pairwise parameter estimates for $\delta=1$ and $\delta=5$ for a single realization of the Brown--Resnick process. The pairwise estimates are significantly worse in terms of variance and bias across the parameter space than the estimates for $\delta = 2$ displayed in Figure~\ref{fig:brparameters}. Figures~\ref{fig:brfiveparamsdelta1} and \ref{fig:brfiveparamsdelta5} show the analogous results for 5 i.i.d.\ realizations of the Brown--Resnick process. While the variance of the pairwise estimates decreases in the case of multiple realizations, the estimates retain the same qualitative patterns as in the single realization case and are overall less accurate than the pairwise estimates for the multiple realization case when $\delta = 2$ shown in Figure~\ref{fig:brfiveparams}.

\begin{figure}[t!]
    \centering
    \includegraphics[scale = .15]{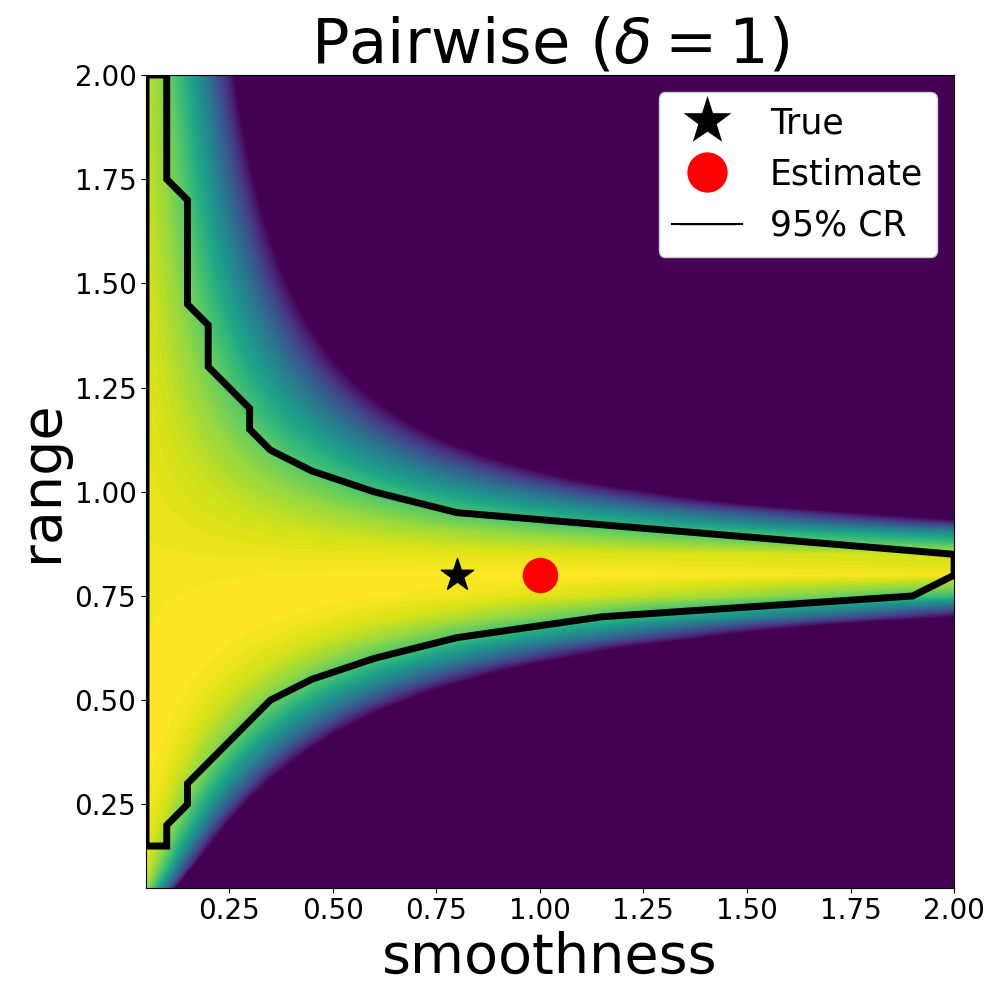}
    \includegraphics[scale = .15]{visualizations/confidence_regions/surfaces/BR/25_by_25/pairwise/dist_2/unadjusted/10_pwl_confidence_region_95_smooth_0.8_range_0.8_rep_14}
    \includegraphics[scale = .15]{visualizations/confidence_regions/surfaces/BR/25_by_25/neural/uncalibrated/10_uncalibrated_neural_confidence_region_95_smooth_0.8_range_0.8_rep_14}
    \includegraphics[scale = .15]{visualizations/confidence_regions/surfaces/BR/25_by_25/neural/calibrated/10_calibrated_neural_confidence_region_95_smooth_0.8_range_0.8_rep_14}
\caption[short]{The unadjusted pairwise likelihood surfaces for distance cut-off $\delta=1$ (far left) and $\delta = 2$ (center left) and neural likelihood surface before calibration (center right) and after calibration (far right) for a realization of a Brown--Resnick process with parameters $\nu = 0.8$ and $\lambda = 0.8$. In each figure, the color scale ranges from the maximum value of the surface to ten units less than the maximum value.}
\label{fig:additionalbrsurfaces}
\end{figure}

We compare the unadjusted pairwise likelihood surfaces for $\delta = 1,2$ to the uncalibrated and calibrated neural likelihood surfaces in Figure~\ref{fig:additionalbrsurfaces}. For $\delta = 1$, the surface is highly uninformative because the area of high likelihood is large. As $\delta$ increases, the area of high likelihood decreases and the surface can become more informative, yet more biased, as shown by the surface for $\delta = 2$. We do not display the unadjusted pairwise likelihood surface for $\delta = 5$ because the area of high likelihood is so sharply concentrated that it carries little meaningful information about the uncertainty of the parameter.

\end{document}